%% file: main.tex
\documentclass[10pt,journal,compsoc]{IEEEtran}

\ifCLASSOPTIONcompsoc
  \usepackage[nocompress]{cite}
\else
  \usepackage{cite}
\fi

\ifCLASSINFOpdf
\else
\fi

\usepackage{algorithmic}
\usepackage{algorithm}
\usepackage{multirow}
\usepackage{color}
\usepackage{comment}
\usepackage{graphicx}

\newcommand{\app}{APIRec-CST}
\newcommand{\APINetwork}{API Context Graph Network}
\newcommand{\TokenNetwork}{Code Token Network}
\newcommand{\singleModel}{APIRec-SO}
\newcommand{\GraLan}{GraLan}

\newcommand{\TreeLSTM}{Tree-LSTM}

\begin{document}
%
\title{Holistic Combination of Structural and Textual Code Information for Context based API Recommendation}
%
%
%
%

\author{Chi~Chen,
        Xin~Peng,~\IEEEmembership{Member,~IEEE,}
        Zhenchang~Xing,
        Jun~Sun,
        Xin~Wang,
        Yifan~Zhao,
        and~Wenyun Zhao
         
\IEEEcompsocitemizethanks{
\IEEEcompsocthanksitem X. Peng is the corresponding author.
\IEEEcompsocthanksitem C. Chen, X. Peng, X. Wang, Y. Zhao and W. Zhao are with the School of Computer Science and the Shanghai Key Laboratory of Data Science, Fudan University, Shanghai, China, and Shanghai Institute of Intelligent Electronics \& Systems, China.
\IEEEcompsocthanksitem Z. Xing is with the Australian National University, Australia.
\IEEEcompsocthanksitem J. Sun is with the Singapore Management University, Singapore.
\protect\\
}
\thanks{}
}

\IEEEtitleabstractindextext{%
\begin{abstract}
Context based API recommendation is an important way to help developers find the needed APIs effectively and efficiently.
For effective API recommendation, we need not only a joint view of both structural and textual code information, but also a holistic view of correlated API usage in control and data flow graph as a whole.
Unfortunately, existing API recommendation methods exploit structural or textual code information separately.
In this work, we propose a novel API recommendation approach called \app~(API Recommendation by Combining Structural and Textual code information). 
\app~is a deep learning model that combines the API usage with the text information in the source code based on an API Context Graph Network and a Code Token Network that simultaneously learn structural and textual features for API recommendation. 
We apply \app~to train a model for JDK library based on 1,914 open-source Java projects and evaluate the accuracy and MRR (Mean Reciprocal Rank) of API recommendation with another 6 open-source projects.
The results show that our approach achieves respectively a top-1, top-5, top-10 accuracy and MRR of 60.3\%, 81.5\%, 87.7\% and 69.4\%, and significantly outperforms an existing graph-based statistical approach and a tree-based deep learning approach for API recommendation.
A further analysis shows that textual code information makes sense and improves the accuracy and MRR.
We also conduct a user study in which two groups of students are asked to finish 6 programming tasks with or without our \app~plugin.
The results show that \app~can help the students to finish the tasks faster and more accurately and the feedback on the usability is overwhelmingly positive.
\end{abstract}

\begin{IEEEkeywords}
API, recommendation, deep learning, data flow, control flow, text
\end{IEEEkeywords}}

\maketitle

\IEEEdisplaynontitleabstractindextext

%
\IEEEpeerreviewmaketitle

\IEEEraisesectionheading{\section{Introduction}\label{sec:introduction}}
\input{introduction}


 

\section{Motivation}\label{sec:motivation}

\input{motivation}

\section{Background}\label{sec:background}
\input{backgroud}

\section{Approach}
\input{approach}

\section{Evaluation}
\input{evaluation}

\section{Related Work}
\input{related_work}

\section{Conclusion}
\input{conclusion}


\ifCLASSOPTIONcaptionsoff
  \newpage
\fi

\bibliographystyle{IEEEtran}
\bibliography{codeRec}

\end{document}

%% file: introduction.tex
\IEEEPARstart{I}{n} modern software development, developers heavily rely on APIs (Application Programming Interfaces).
When developers do not know which API(s) to use for a desired feature, automatic API recommendation is an important way to help developers find the needed APIs effectively and efficiently.
In general, API recommendation methods learn explicit or implicit API usage patterns from a large code base and then match partially written code with the patterns to recommend APIs.
Existing methods differ in the types of code information they model and how they model code information.

Source code contains two core types of information: structural and textual.
Structural code information, such as control and data flow, represents program logic which can be captured using a graph representation;
textual code information, such as code comments, method names, variable names, reflects the semantics of the code in natural language.
Take the code snippet in Fig.~\ref{fig:get hash code from file} as an example.
Note that the correct API statement at line 8 should be $hashCode=str.hashCode()$.
The method name ``computeHashCode'' and the variable name ``hashCode'' reflect the intent of this method (assuming the proper tokenization of these names).
The method body uses multiple APIs which implement three pieces of correlated program logics: 1) use a reader to read contents from a file line by line (line 3/4/5/6/11/12); 2) compute the hash code of the content (line 8); 3) add the hash value into a created list (line 2/7/9).
These program logics can be modeled in a control and data flow graph as shown in Fig.~\ref{fig:API context graph}.
Note that variable names (e.g., ``path'', ``result'', ``rd'', ``br'', ``str'', ``hashCode'') are helpful for the understanding of relevant structural program logics.

For effective API recommendation, we need not only a joint view of both structural and textual code information, but also a holistic view of correlated API usage in control and data flow graph as a whole.
Unfortunately, existing API recommendation methods exploit structural or textual code information separately.
Based on the observation of linguistic naturalness of source code~\cite{ICSE12Naturalness}, many approaches~\cite{ICSE12Naturalness,MSR13tatisticalLearninig,FSE13SLAMC,FSE14Localness} have been proposed that rely on statistical language models for code auto-completion and API recommendation.
The adopted statistical language models can be simple or enhanced n-gram model~\cite{ICSE12Naturalness,MSR13tatisticalLearninig,FSE13SLAMC,FSE14Localness} or complex deep learning models (e.g., Recurrent Neural Network (RNN))~\cite{PLDI14SLANG,LSTMForCode,SANER18Dnn4C}.
No matter which types of statistical language models to use, these approaches treat code as a sequence of text tokens (which may sometimes be enriched with simple syntactic information such as program construct keywords and data types), but do not exploit structural code information of source code.
As such, they cannot properly model the long-range dependencies between correlated but far-away API usage due to the limitation of the length of a sequence.

To overcome the limitation of token-sequence-based API recommendation, another important line of API recommendation methods~\cite{ICSE15GraLan,ASE18RecRank} analyze control and data flow graph for recommending APIs.
However, these methods usually base their recommendation on the enumeration of control and data flow subgraphs, but lack a holistic view of the overall program logic.
Consider the code snippet in Fig.~\ref{fig:get hash code from file}.
Fig.~\ref{fig:subgraphs} shows nine control-and-data-flow subgraphs for this code snippet.
Assume developers do not know the ``java.lang.String.hashCode'' API to be used at line 8.
Unfortunately, existing methods recommend ``java.io.BufferedReader.readLine'' based on the fourth subgraph in Fig.~\ref{fig:subgraphs} or ``while'' based on the sixth subgraph.
Different subgraphs are treated independently for recommending relevant APIs.
As smaller subgraphs usually appear more frequently than larger subgraphs, APIs from smaller subgraphs that capture only a partial aspect of the overall program logic often overshadow APIs from larger subgraphs that capture more holistic view of the program logic.


In this work, we propose a novel API recommendation approach called \app~(\textbf{API} \textbf{Rec}ommendation by \textbf{C}ombining \textbf{S}tructural and \textbf{T}extual code information), which addresses the limitation of independent modeling of structural and textual code information and the lack of holistic reasoning of code structure in existing API recommendation approaches.
\app~is a deep learning model that combines the API usage with the text information in the source code based on an \APINetwork~and a \TokenNetwork.
As such, it can simultaneously learn structural and textual features for API recommendation.
\app~uses an API context graph to model API usage in a control and data flow graph for the entire method, rather than independent partial subgraphs as  in existing methods~\cite{ICSE15GraLan}.
Our API context graph contains the holistic semantics of the API usage in the source code around the location for API recommendation.
From this API context graph, the \APINetwork~learns to extract informative structural features for API recommendation.
The textual code information in the source code, such as method names, parameter names and variable names, is processed as a bag of code tokens which is fed into the \TokenNetwork~to infer the developer's intent jointly with the \APINetwork.

We conduct a series of experiments to evaluate the effectiveness of \app.
Our results show that \app~significantly outperforms an existing graph-based statistical approach and a tree-based deep learning approach for API recommendation.
The overall top-1 accuracy of \app~is about 60.3\%, the top-5 accuracy is about 81.5\%, the top-10 accuracy is about 87.7\% and the MRR is about 69.4\%.
In addition, our analysis shows that textual code information makes sense and improves the accuracy and MRR.
The results of our user study with 18 students and 6 programming tasks show that \app~can help the students finish the tasks faster and more accurately and the feedback on our tool's usability is overwhelmingly positive.

The main contributions of this work are as follows:
\begin{itemize}
	\item We propose an API recommendation approach called \app~that combines structural and textual code information in the source code by jointly learning a graph-based deep learning model and a token-based deep learning model for effective API recommendation.

	\item We implement \app~as a tool that supports the efficient model training and API inference with GPU acceleration.
	
	\item We evaluate the effectiveness of \app~for recommending APIs with both automatically constructed test instances and real programming tasks.
\end{itemize}

%% file: motivation.tex
\begin{figure}
	\centering
	\includegraphics[scale=0.13]{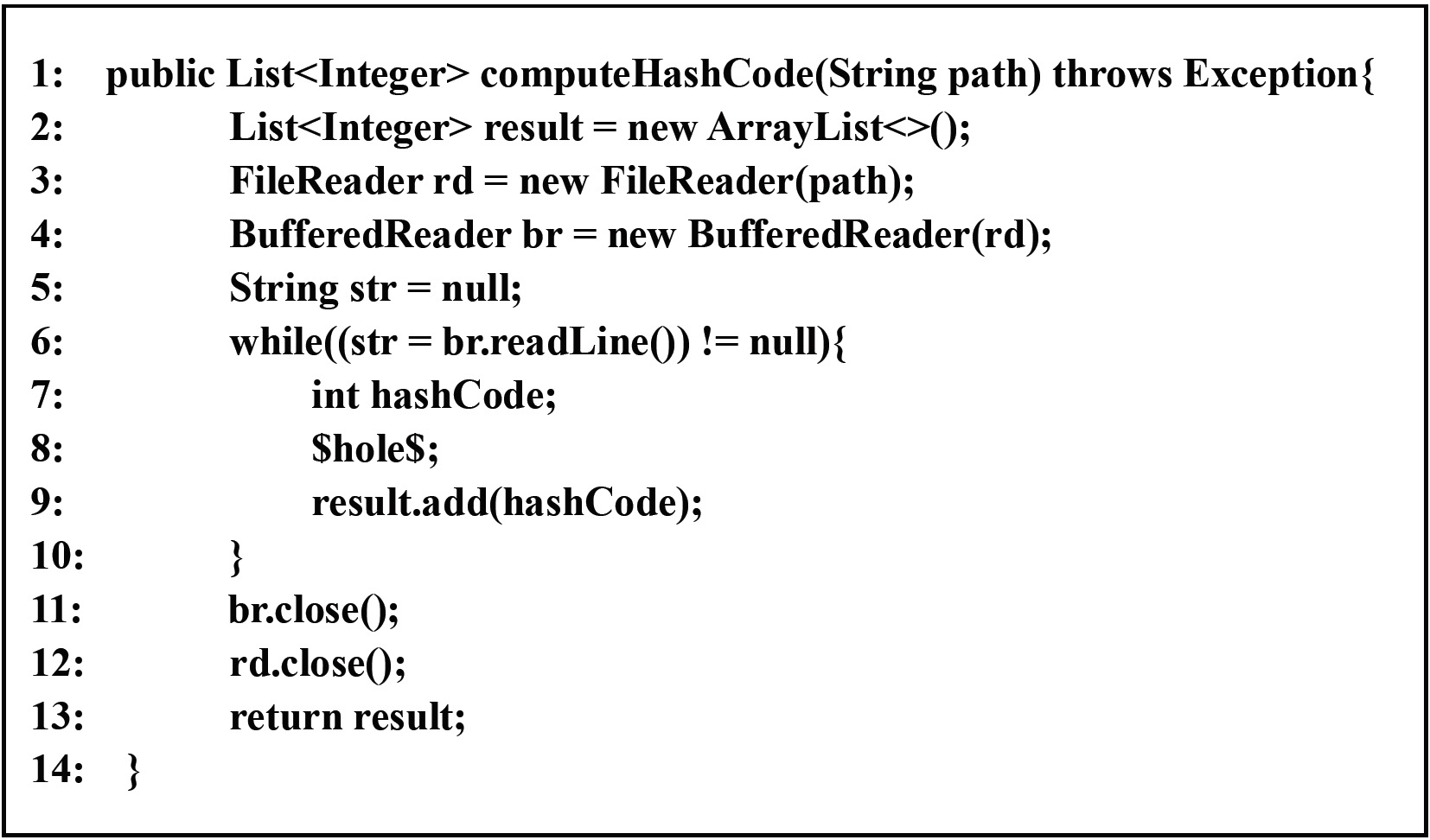}
	\caption{Example of Computing HashCode of Content from File}
	\label{fig:get hash code from file}
\end{figure}
\begin{figure}
	\centering
	\includegraphics[scale=0.13]{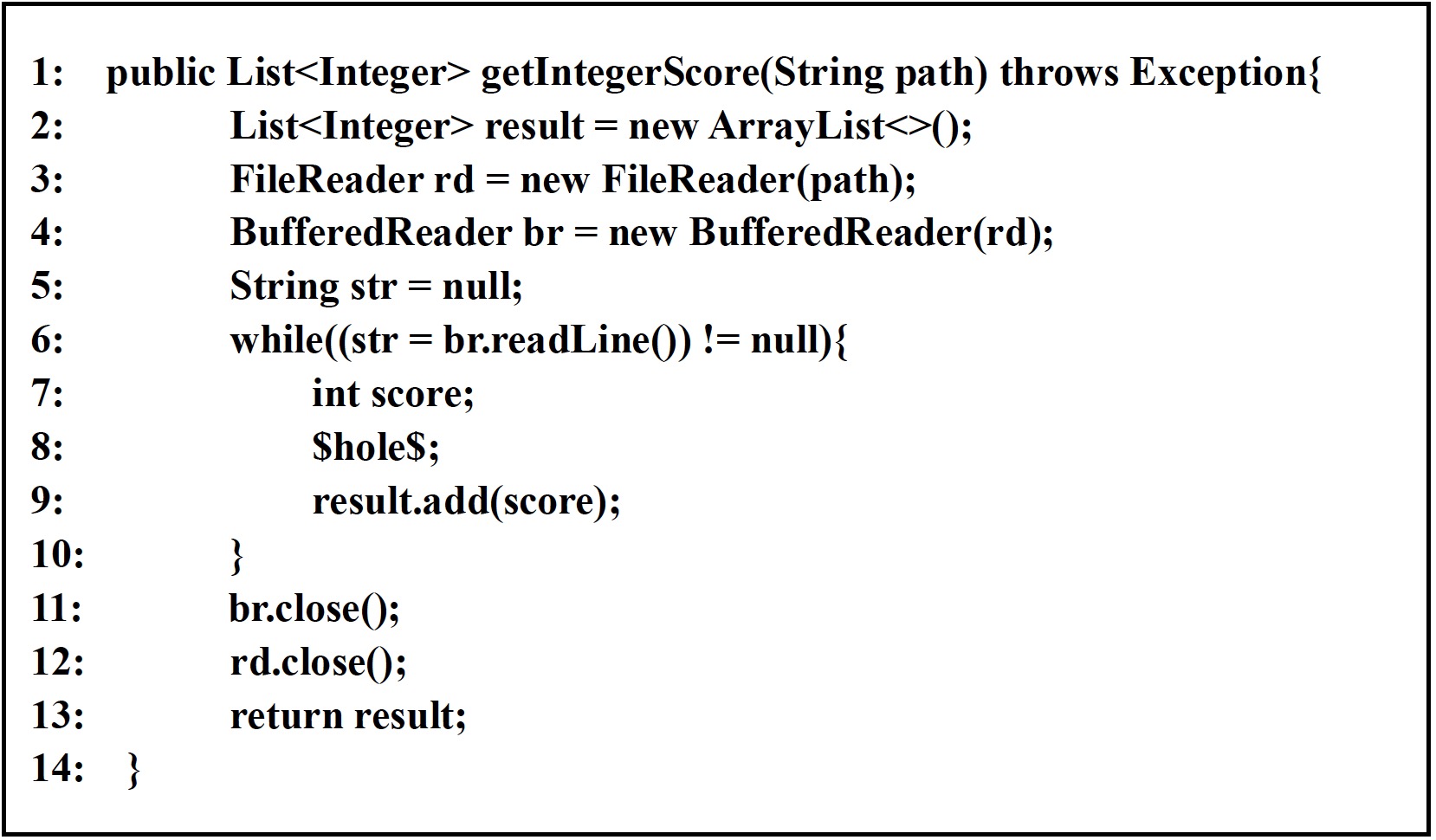}
	\caption{Example of Getting Integer Score from File}
	\label{fig:get score from file}
\end{figure}

We use the code examples in Fig.~\ref{fig:get hash code from file} and Fig.~\ref{fig:get score from file} to motivate the need for holistic combination of structural and textual code information for API recommendation.
The example in Fig.~\ref{fig:get hash code from file} is to implement a method to compute the hash code of the content from a file line by line and then adds the computed hash code into a list.
The developer has written the code he knows and needs help to complete the remaining code.
The line marked as \emph{hole} is the location that the developer requests the recommendation of proper APIs for computing the hash code of the content of a string.

\begin{figure}
    \centering
    \includegraphics[scale=0.18]{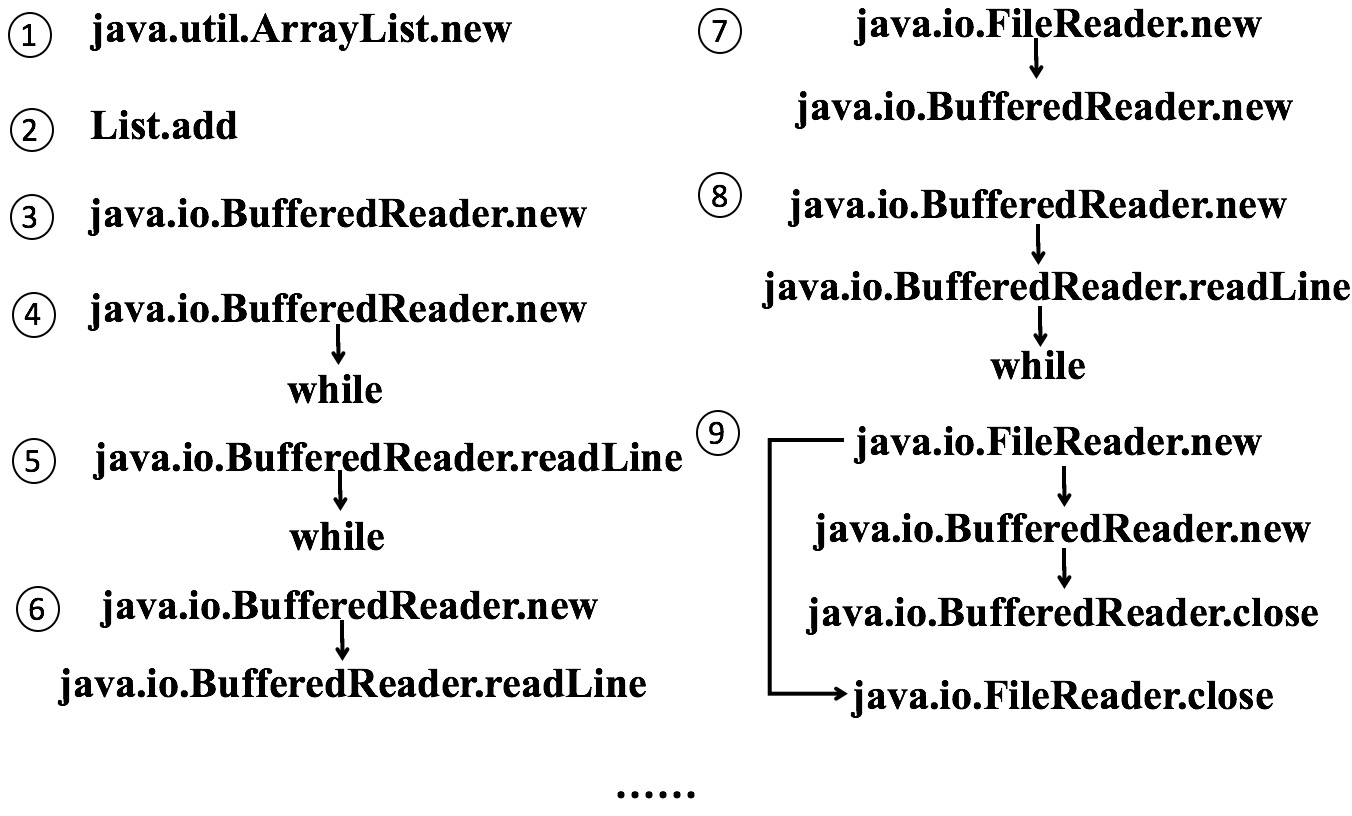}
    \caption{Control-and-Data-Flow Subgraphs of the Code in Fig.~\ref{fig:get hash code from file}}
    \label{fig:subgraphs}
\end{figure}

We can see that this program contains rich structural code information (i.e., multiple APIs and control and data flow among these APIs).
We can get many subgraphs of different sizes according to control and data flow, such as the nine subgraphs shown in Fig.~\ref{fig:subgraphs}.
Note that each subgraph is labeled with a serial number for the convenience of discussion.
We do not list all the subgraphs for the code in Fig.~\ref{fig:get hash code from file} due to the space limitation.
As we can see, each subgraph reflects partial program logic (semantics).
For example, the seventh subgraph reflects the semantics of creating readers for reading a file.
As another example, the fifth subgraph reflects the semantics of reading contents line by line.
None of the subgraphs (including those not listed in the paper) independently can reflect the expected semantics (i.e., computing the hash code of a string) at the location of \emph{hole}.

If the developer uses existing tools such as \GraLan~\cite{ICSE15GraLan} that recommends APIs based on such subgraphs, he cannot get the correct API recommendation.
Table~\ref{tab:recommendations} lists the top-10 recommendations by \GraLan.
The first column is the ranking of each recommendation.
The second column lists the ten recommendations.
The third column is the serial number of the subgraphs in Fig.~\ref{fig:subgraphs} used as the parent graph based on which the corresponding recommendation is generated.
In \GraLan, each subgraph is considered as a context parent graph to generate child graphs (each child graph has one more node than its parent graph and the extra node is considered as a candidate API recommendation).
From Table~\ref{tab:recommendations}, we can see that the top-10 recommendations by \GraLan~are generated based on partial program semantics and thus miss the correct recommendation.

In order to recommend the correct API, we need a holistic view of the overall program logic in the entire method.
Hence, we represent the API usage in a whole control and data flow graph called API context graph (as shown in Fig.~\ref{fig:API context graph}) instead of subgraphs for the entire method.
The API context graph not only contains all semantics in subgraphs, but also integrates these semantics as a whole.
The details of how to construct an API context graph will be introduced in Section~\ref{sec:representation}.
From the API context graph, we can see that it contains the following two major semantics: semantics-1) use a reader to read contents from a file line by line; semantics-2) add a value into a created list. Since these semantics are in one entire graph, they can be integrated to infer the semantics at the hole.

\begin{table}[t]
\scriptsize
\centering
\caption{Top-10 Recommendations by \GraLan~\cite{ICSE15GraLan} for the Code Snippet in Fig.~\ref{fig:get hash code from file}}
\label{tab:recommendations}
\begin{tabular}{|c|c|c|}
\hline
\textbf{Rank} & \textbf{Recommendation} & \textbf{Parent Graph} \\
\hline
1 & java.util.List.add & 1 \\ 
2    & java.util.ArrayList.new & 2 \\
3    & java.io.BufferedReader.readLine & 4 \\
4    & java.io.BufferedReader.new & 5\\
5    & while & 6 \\
6    & java.io.BufferedReader.close & 8 \\
7    & if & 2 \\
8    & for & 2 \\
9    & java.util.ArrayList.add & 1 \\
10    & java.io.InputStreamReader.new & 3 \\
\hline
\end{tabular}
\end{table}

When observing these two semantics in a holistic view, we can find that the declared $String$ variable ``str'' is just used to store the content from the file but not used any more in semantics-1.
Furthermore, the declared $int$ variable ``hashCode'' is not assigned a value in semantics-2.
In addition, there lack of APIs to connect semantics-1 and semantics-2 to make the program logic complete.
From this holistic view, we can infer that the semantics at the \emph{hole} is to get a value of $int$ type based on some kind of processing of a variable of $String$ type.
Note that the subgraph can be a whole graph in \GraLan, but the larger a graph is, the less frequent it may occur in the training data which may cause the data sparsity issue.
Our deep learning model learns a vector representation for each entire graph based on an information diffusion mechanism of all nodes and edges.
In this way, each entire graph that has a distinct semantics will have a meaningful vector representation, no matter how large the graph is and how frequent the graph occurs in the code base.
As such, our model does not suffer from the data sparsity issue.

However, we still cannot recommend the exact API needed at the $hole$ in Fig.~\ref{fig:get hash code from file}, if we just consider the structural code information in this example.
This is because we cannot decide what kind of processing should be performed on the variable of $String$ type.
Let us see the code snippet in Fig.~\ref{fig:get score from file}.
The developer needs to implement a method to read scores stored in a file, convert each score to an integer and add it into a list for further use.
We can see that the code in Fig.~\ref{fig:get score from file} is structurally very similar to the code in Fig.~\ref{fig:get hash code from file}, because the program logics for reading file and list addition are the same.
The API context graph of the code in Fig.~\ref{fig:get score from file} is the same as that of the code in Fig.~\ref{fig:get hash code from file}, but the expected APIs at \emph{hole} are different.
If the developer requests API recommendation for these two code snippets, we should distinguish the different intents in the two code snippets.
To that end, textual code information in code becomes very useful for inferring code intents.
In Fig.~\ref{fig:get hash code from file}, the method name ``computeHashCode'' and variable name ``hashCode'' imply that the processing on the variable of $String$ type is likely relevant to hash code processing.
In Fig.~\ref{fig:get score from file}, the method name ``getIntegerScore'' and variable name ``score'' can imply that the processing on the variable of $String$ type is likely relevant to String-Integer conversion.

To sum up, a joint view of both structural and textual code information and a holistic view of correlated API usage in control and data flow graph of the entire method is required for effective API recommendation.


%% file: backgroud.tex
\begin{figure}
    \centering
    \includegraphics[scale=0.4]{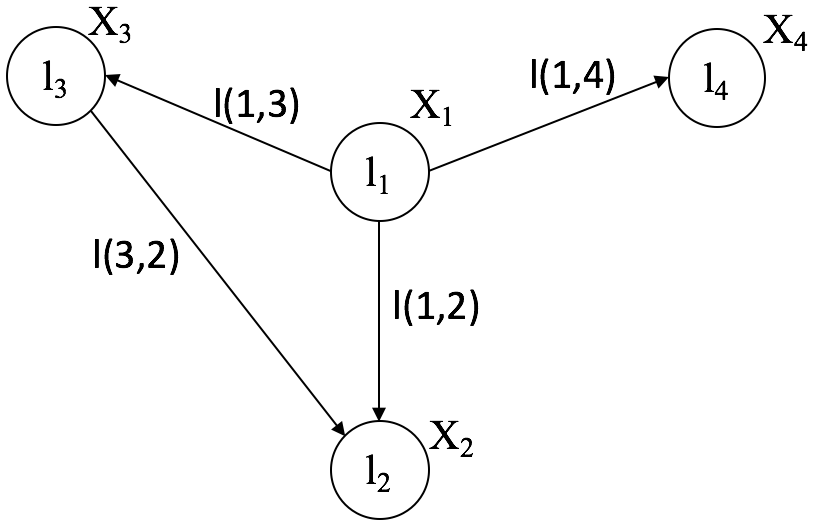}
    \caption{Graph Example}
    \label{fig:graph example}
\end{figure}

In this work, we adopt Graph Neural Networks (GNNs), in particular, Gated Graph Neural Networks (GG-NNs)~\cite{GGNN}, for API recommendation. An API usage can be naturally represented in the form of a graph where the nodes represent APIs and edges represent control/data flow between nodes. Furthermore, the nodes and edges can be labeled with additional context information, e.g., the nodes can be labeled with API calls and the edge labels can be used to distinguish control flow and data flow.

%

GNNs are a neural network model which take graph structures as inputs. GNNs are based on an information diffusion mechanism and work effectively for a variety of graphs, e.g., directed or undirected graphs and cyclic or acyclic graphs. In GNNs, each node of the graph corresponds to a unit. The unit captures the current state of a node and is used to compute the next state of the node when activated. The units update their states and exchange information until they reach a stable equilibrium~\cite{GNN}.
The state of a node is composed of the label of the node, the labels of its incoming and outgoing edges and the states and labels of neighbor nodes with a parametric function.
Formally, a state \textbf{x$_{n}$}(t) at the $t$th iteration of a node n is defined as follows~\cite{GNN}.
\begin{equation}
\textbf{x$_{n}$}(t)=\textbf{f$_{w}$}(\textbf{l$_{n}$},\textbf{l$_{co[n]}$},\textbf{x$_{ne[n]}$}(t-1),\textbf{l$_{ne[n]}$}),
\label{equ:state}
\end{equation}
where \textbf{f$_{w}$} is a parametric function, \textbf{l$_{n}$} is the label of node n, \textbf{l$_{co[n]}$} are the labels of edges containing node n, \textbf{x$_{ne[n]}$}(t-1) are the states of nodes in the neighborhood of node n at the $(t$$-$$1)$th iteration, and \textbf{l$_{ne[n]}$} are the labels of nodes in the neighborhood of node n.
In this way, each node can get a node representation.
Take the graph in Figure~\ref{fig:graph example} as an example.
The state \textbf{x$_{1}$} of node 1 at time t is computed as \textbf{x$_{1}$}(t)=\textbf{f$_{w}$}(\textbf{l$_{1}$},\textbf{l$_{(1,2)}$,l$_{(1,3)}$,l$_{(1,4)}$},\textbf{x$_{2}$}(t-1),\textbf{x$_{3}$}(t-1),\textbf{x$_{4}$}(t-1),\textbf{l$_{2}$,l$_{3}$,l$_{4}$}), where \textbf{l$_{1}$} is the label of node 1, \textbf{l$_{(1,2)}$, l$_{(1,3)}$, l$_{(1,4)}$} are the labels of edges connected with node 1, \textbf{x$_{2}$}(t-1), \textbf{x$_{3}$}(t-1), \textbf{x$_{4}$}(t-1) are the states of the neighboring nodes (i.e., node 2, node 3 and node 4) of node 1 at time t-1 and \textbf{l$_{(1,2)}$, l$_{(1,3)}$, l$_{(1,4)}$} are the labels of these neighbors of node 1. The state of a node is connected with other nodes in the graph as nodes can communicate with each other based on the information diffusion mechanism.
Through training, GNNs can be applied for subgraph matching, mutagenesis, and web page ranking~\cite{GNN}.



GG-NNs~\cite{GGNN} are based on GNN. The difference is that GNNs apply Almeida-Pineda algorithm~\cite{Almeida,Pineda} for computing gradients, whereas GG-NNs apply back-propagation through time with Gated Recurrent Units~\cite{GRU} for computing gradients.
GG-NNs use a soft attention mechanism to decide which nodes are more relevant to compute the final vector representation of the graph. The graph level representation vector \textbf{x$_{g}$} is computed as follows~\cite{GGNN}.
\begin{equation}
\textbf{x$_{g}$}=tanh\Bigg(\sum_{n \in N}\sigma\Big(i(\textbf{x$_{n}$}(t),\textbf{l$_{n}$})\Big) \odot tanh\Big(j(\textbf{x$_{n}$}(t),\textbf{l$_{n}$})\Big)\Bigg),
\label{equ:soft_attention}
\end{equation}
where $\sigma$(i(\textbf{x$_{n}$}(t),\textbf{l$_{n}$})) works as a soft attention mechanism, $i$ and $j$ are neural networks taking as input the concatenation of \textbf{x$_{n}$}(t) and \textbf{l$_{n}$} and output real-valued vectors~\cite{GGNN}, and $\odot$ is element-wise multiplication.

To get a graph representation, GNNs require creating a dummy super node which is connected to all other nodes by a special type of edge~\cite{GGNN}.
Doing so in our context may destroy the structural code information of the source code itself. 
In addition, the soft attention mechanism of GG-NNs can help us to identify which nodes (i.e., APIs) in the API context graph are more important for API recommendation.
In GG-NNs, the final representation of a graph is the accumulated information of each node with its importance computed through the soft attention mechanism. In this way, the final representation of a graph is a a holistic representation of all nodes.
Therefore, we choose GG-NNs as our deep neural networks to learn the features of API context graphs from a holistic view. More details of GNNs and GG-NNs can be referred to~\cite{GNN,GGNN}.


%% file: approach.tex
In this section, we present the detailed design of \app.
It takes a program with a hole as input, and outputs a ranked list of API recommendations for filling the hole.

\subsection{Program Representation}\label{sec:representation}
\input{representation}

\subsection{Architecture}
\input{architecture}

\subsection{Training Corpus Construction}
\input{corpus}

%% file: representation.tex
Given a program with a hole, \app~first constructs an API context graph and a bag of code tokens. The API context graph is a graph representation of structural code information of the user-provided program, whereas the code tokens (including the method name, parameter names and variable names) capture the textual code information.

\noindent \emph{An API context graph} is a directed graph $(N, E)$ where $N$ is a set of nodes and $E \subseteq N \times N$ is a set of edges. Each node in $N$ represents an API method call, an API field access, a variable declaration, an assignment, a control unit or a hole. Furthermore, each node is labeled differently according to its type. Table~\ref{table:representation} shows how each type of node is labeled.
We use a special node labeled with $Hole$ (called hole node hereafter) to represent the hole. 
There is an edge $(n, n') \in E$ if and only if one of the following conditions is satisfied.
\begin{itemize}
    \item There is a direct control flow from $n$ to $n'$;
    \item There is a direct data flow from $n$ to $n'$;
    \item $n'$ is the hole node and $n$ is a node representing the preceding statement in the program or $n$ is the hole node and $n'$ is a node representing the subsequent statement in the program. 
\end{itemize}
Given an edge $(n, n')$, we say that $n$ is the parent node of $n'$ and $n'$ is the child node of $n$. In \app, the edges in an API context graph are distinguished by labeling them with different types, i.e., an edge is labeled \emph{control flow} (Type c) is there is direct control flow and no direct data flow; an edge is labeled \emph{data flow} (Type d) is there is direct data flow and no direct control flow; an edge is labeled \emph{control and data flow} (Type cd) is there are both direct control flow and direct data flow; and an edge is labeled \emph{special flow} (Type s) if its source node or target node is the hole.
Note that the \emph{special flow} edge makes sure that the hole node is connected to its context.


Given a program, \app~systematically builds the API context graph statically. First, \app~builds the AST (i.e., Abstract Syntax Tree) of the program.
Then it creates nodes and edges in the API context graph for each statement in the program based on the AST in the following way.
\begin{itemize}
\item If the statement is an API method call, an API field access, a variable declaration or an assignment, a node is created according to the corresponding node type in Table~\ref{table:representation}.
 Note that if the parameter of an API method call is also an API method call or an API field access, \app~first creates a node for the parameter.
\item If the current statement is an expression that includes several API method calls or API field accesses, \app~creates a node for each API method call or API field access one by one. 
\item If the current statement is a control statement, \app~creates a node for the control unit according to its type and several other nodes together with edges connecting them as shown in Table~\ref{table:control statement rules}. 
    For example, if the current statement is a \emph{while} statement, \app~first creates a \emph{While} node, a  \emph{Condition} node, and a  \emph{Body} node.
    Two Type c edges are introduced, one from the \emph{While} node to the \emph{Condition} node and the other from the \emph{While} node to the \emph{Body} node.
\end{itemize}

\begin{table*}[t]
  \scriptsize
  \centering
  \caption{Labels of Different Types of Nodes in API Context Graphs}
  \label{table:representation}
  \begin{tabular}{|l|l|l|}
    \hline
    \textbf{Node Type} &
    \textbf{Label} &
    \textbf{Example} \\
    \hline
    Vari. Decl. &
    [Full Class Name].Declaration &
    \texttt{String str;} $\rightarrow$ java.lang.String.Declaration \\
    \hline
    Vari. Decl. with Constant Assignment &
    [Full Class Name].Constant &
    \texttt{String str = "str";} $\rightarrow$ java.lang.String.Constant \\
    \hline
    Vari. Decl. with Null Assignment &
    [Full Class Name].Null &
    \texttt{String str = null;} $\rightarrow$ java.lang.String.Null \\
    \hline
    Vari. Decl. with Object Creation &
    [Full Class Name].new([Parameter Types]) &
    \texttt{File file = new File(path);} $\rightarrow$ java.io.File.new(java.lang.String) \\
    \hline
    API Method Call &
    [Full Method Name]([Parameter Types])&
    \texttt{builder.append("str");} $\rightarrow$ java.lang.StringBuilder.append(java.lang.String) \\
    \hline
    API Field Access &
    [Full Field Name] &
   \texttt{System.out;} $\rightarrow$ java.lang.System.out\\
   \hline
    \multirow{4}{*}{\shortstack{Cascading API Method Call \\(API Field Access)}}& [Full Method Name]([Parameter Types])& \texttt{builder.append("str").toString(); $\rightarrow$}\\
     &.[Method Name]([Parameter Types])& java.lang.StringBuilder.append(java.lang.String).toString()\\
    \cline{2-3} & [Full Field Name] & \texttt{System.out.println("str"); $\rightarrow$} \\
     &.[Method Name]([Parameter Types]) & java.lang.System.out.println(java.lang.String)\\
    \hline
    \multirow{4}{*}{\shortstack{Nested API Method Call \\ (API Field Access)}}& [Full Method Name]([Parameter Types])& \texttt{writer.write(sb.toString());} $\rightarrow$ java.lang.StringBuilder.toString()\\
     &[Full Method Name]([Parameter Types])& java.io.FileWriter.write(java.lang.String)\\ 
    \cline{2-3} & [Full Field Name] & \texttt{label.setForeground(Color.blue);} $\rightarrow$ java.awt.Color.blue\\
     &[Full Method Name]([Parameter Types]) & javax.swing.JLabel.setForeground(java.awt.Color)\\
    \hline
    Control Unit &
    [Control Unit Name]&
    \texttt{if} $\rightarrow$ If\\
    \hline
  \end{tabular}
\end{table*}

\begin{table*}[t]
  \scriptsize
  \centering
  \caption{API Context Graph Nodes and Edges for Control Statements}
  \label{table:control statement rules}
  \begin{tabular}{|p{2.5cm}|p{2.2cm}|p{8.5cm}|}
    \hline
    \textbf{Control Statement Type} &
    \textbf{Node of Control Unit} &
    \textbf{Nodes and Edges} \\
    \hline
    \multirow{3}{*}{\emph{if} statement} & \multirow{3}{*}{\textbf{If}} & a \textbf{Condition} node and a Type c edge from \textbf{If} node to \textbf{Condition} node\\
 \cline{3-3}
 & & a \textbf{Then} node and a Type c edge from \textbf{If} node to \textbf{Then} node\\
 \cline{3-3}
 & & a \textbf{ElseIf/Else} node and a Type c edge from \textbf{If} node to \textbf{ElseIf/Else} node\\
 \hline
     \multirow{2}{3cm}{\emph{while/do\\for/foreach} statement} & \multirow{2}{3cm}{\textbf{While/DoWhile\\For/Foreach}} & a \textbf{Condition} node and a Type c edge from \textbf{While/DoWhile/For/Foreach} node to \textbf{Condition} node\\
\cline{3-3}
 & & a \textbf{Body} node and a Type c edge from \textbf{While/DoWhile/For/Foreach} node to \textbf{Body} node\\
 \hline
  \multirow{3}{*}{\emph{switch} statement} & \multirow{3}{*}{\textbf{Switch}} & a \textbf{Selector} node and a Type c edge from \textbf{Switch} node to \textbf{Selector} node\\
  \cline{3-3}
       & & a series of \textbf{Case} nodes and Type c edges from \textbf{Switch} node to each \textbf{Case} node\\
  \cline{3-3}
       & & a \textbf{Default} node and a Type c edge from \textbf{Switch} node to \textbf{Default} node\\
  \hline
    \multirow{3}{*}{\emph{try} statement} & \multirow{3}{*}{\textbf{Try}} & a series of \textbf{Catch} nodes and a Type c edge from \textbf{Try} node to the first \textbf{Catch} node\\
    \cline{3-3}
       & & Type c edges connecting \textbf{Catch} nodes in order (such as from first to second, from second to third)\\
    \cline{3-3}
       & & a \textbf{Finally} node and a Type c edge from the last \textbf{Catch} node to \textbf{Finally} node\\
  \hline
  \end{tabular}
\end{table*}

Next, we systematically analyze the control and data dependencies between the nodes (i.e., between the corresponding statements) and introduce the edges accordingly.
Take the \emph{while} statement as an example.
A Type c edge is added from the \emph{Condition} node to the first node created for the condition expression and a Type c edge is added from the \emph{Body} node to the first node created for the loop body. In addition, a Type c edge is added from the \emph{While} node to the first node representing the statement following the loop. 
If the program contains a hole, a Type s edge is added from the node representing the statement preceding the hole to the hole node and a Type s edge is added from the hole node to the node representing the statement succeeding the hole.
Since the control and data dependency analysis is performed statically in~\app, we acknowledge that it might not be fully accurate. 
This is however the standard approach in existing state-of-the-art approaches~\cite{ICSE15GraLan,ASE18RecRank}, since obtaining control/data dependency through dynamic analysis has its limitations as well. Furthermore, since \app~is based on big data, the inaccuracies in individual graph (due to problems like program specific aliasing) are likely filtered out as noises.

\begin{figure}
    \centering
    \includegraphics[scale=0.20]{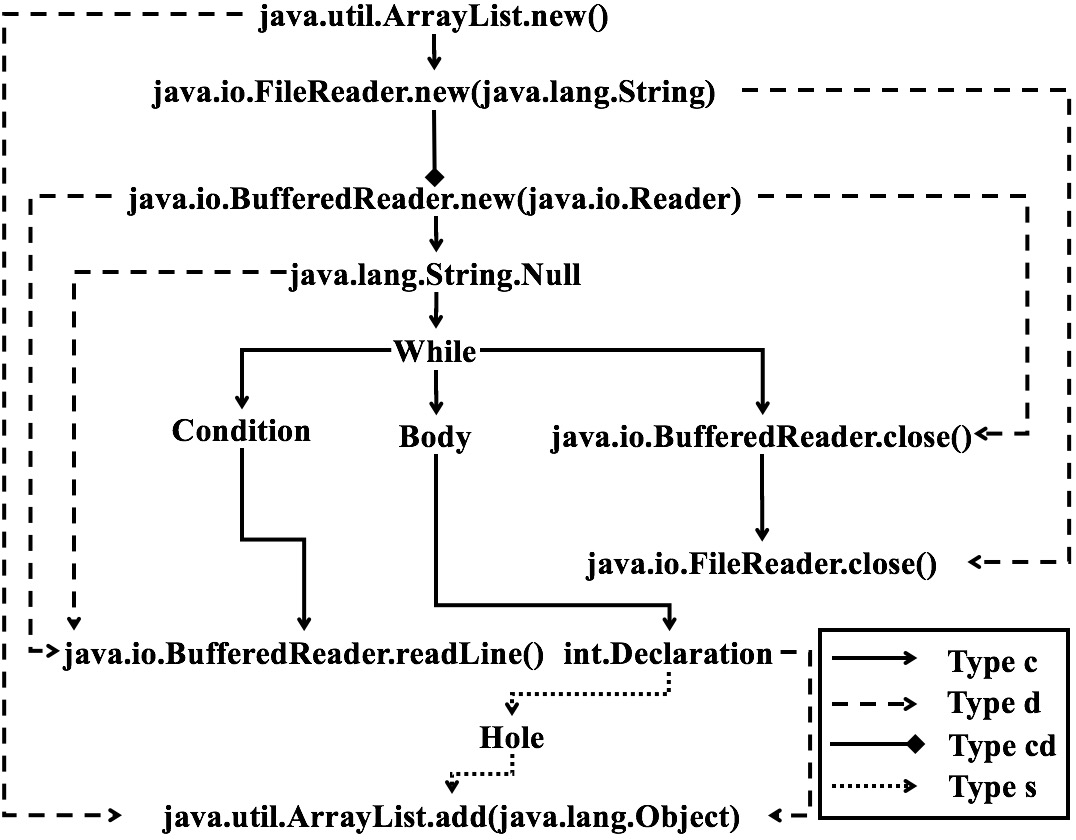}
    \caption{API Context Graph of the Source Code in Fig.~\ref{fig:get hash code from file}}
    \label{fig:API context graph}
\end{figure}

For instance, the API context graph for the program shown in Fig.~\ref{fig:get hash code from file} is shown in Fig.~\ref{fig:API context graph}, where solid lined triangle arrows represent edges labeled with control flow; dashed lined triangle arrows represent edges labeled with data flow; solid lined diamond arrows represent edges labeled with control and data flow; and dotted lined triangle arrows represent edges labeled with special flow. We can see that different from the graph used in \GraLan, each edge is given a type in our API context graph and the structure of our API context graph is closely related to the program structure. In addition, although the program contains a hole, our API context graph is still a connected graph that contains all related structure information, but in \GraLan, a graph is not a connected graph but consists of several context subgraphs around the hole.

\noindent \emph{The bag of code tokens} consists of tokens of the method name, parameter names and variable names.
As mentioned before, it captures the textual code information, which is useful for API recommendation.
The bag of code tokens is collected as follows. First, \app~systematically extracts the method name, parameter names and variable names based on the AST of the program.
Note that \app~only extracts the names of parameters and variables whose types are included in the target library (e.g., JDK).
Second, because developers often use compound or nonstandard word as names, the extracted names are split as tokens.

\app~adopts a simple and efficient rule-based method for splitting names into atomic tokens. First, the numbers in a name are pruned. For example, ``file2'' becomes ``file'' afterwards.
Second, the name is split into multiple tokens using the two special characters ``\_'' and ``\$'' that are often used in naming.
For example, ``file\_name'' is split into ``file'' and ``name''.
Third, each token is further split according to camel case~\cite{camel_case}.
For example, ``fileName'' is split into ``file'' and ``name''.
Next, each token is processed by lemmatization. For example, ``files'' is converted to ``file''.
Lastly, \app~post-processes the tokens by removing duplicated tokens as well as tokens which are meaningless, e.g., one character such as ``i'' and ``j''.
In general, only those tokens which are in the GloVe vocabulary~\cite{glove} are deemed meaningful. The GloVe vocabulary contains 400K unique tokens obtained from Wikipedia and Gigaword.

For instance, the bag of code tokens obtained from the program shown in Fig.~\ref{fig:get hash code from file} includes ``compute'', ``hash'', ``code'', ``path'', ``result'', ``rd'', ``br'' and ``str''.

%% file: architecture.tex
\begin{figure}
    \centering
    \includegraphics[scale=0.23]{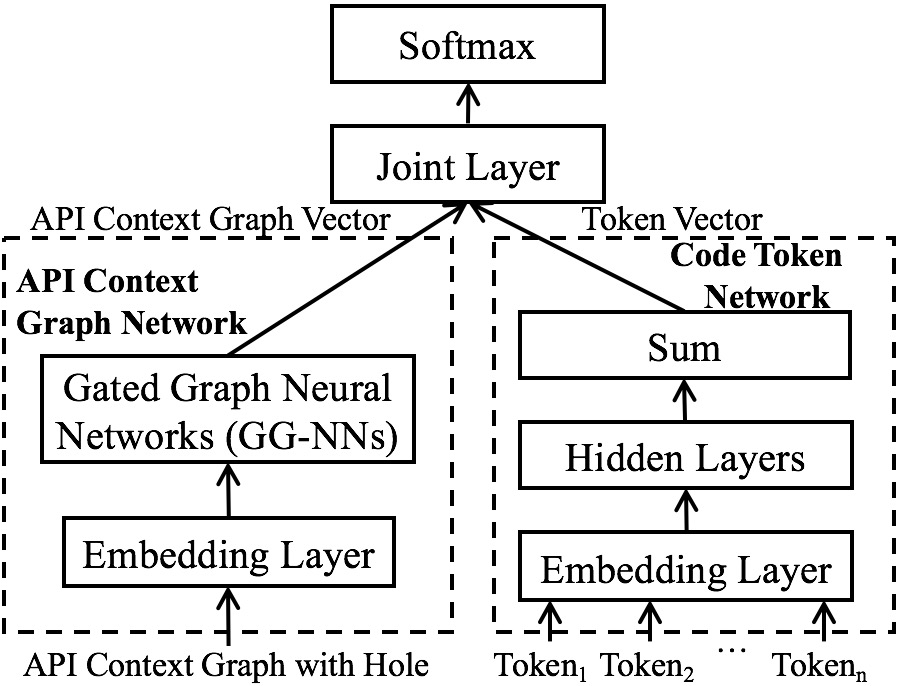}
    \caption{The Overall Architecture of \app}
    \label{fig:architecture}
\end{figure}

Given a program with a hole represented in the form of an API context graph and a bag of code tokens, the task of \app~is to predict what should be for filling the hole. \app~is designed to solve the task based on deep learning techniques. Fig.~\ref{fig:architecture} shows the overall architecture of \app, which consists of two main components i.e., the \APINetwork~and the \TokenNetwork, as well as a joint layer.
The \APINetwork~learns an API context graph vector based on a given API context graph. It consists of an embedding layer and GG-NNs.
The \TokenNetwork~learns a token vector based on a given bag of code tokens. It consists of an embedding layer, multiple hidden layers and a sum operation.
The joint layer is designed to combine the API context graph vector and token vector and output a joint vector.
The softmax function is then used to compute the probabilities of each candidate APIs based on the joint vector.
We introduce each component and the joint layer in the following. \\

\noindent \emph{\APINetwork}
The \APINetwork~takes as input an API context graph (with a hole to be filled) and outputs a vector.
The API context graph is processed as a set of nodes and edges and fed into the network. An embedding layer is first used to embed the node label of each node into an individual vector which is then used as the initial vector of the node annotation in GG-NNs.
Then the nodes and edges are passed into GG-NNs to get an API context graph vector.

In order to get the API context graph vector, GG-NNs first compute the state of each node and the state from the last time step is used as the node representation.
The overall process of computing the state of each node is introduced in Section~\ref{sec:background} and the details can be referred to~\cite{GNN,GGNN}. 
Afterwards the API context graph vector is computed based on the node representations with a soft attention mechanism to decide which nodes are relevant to the current API context graph.
The detailed equation of computing the API context graph vector can be found in Section~\ref{sec:background} and~\cite{GGNN}.
\\

\noindent \emph{\TokenNetwork}
The \TokenNetwork~takes as input the bag of code tokens and outputs a vector. 
To obtain the output token vector, an embedding layer is first used to embed each of the code tokens into an individual vector.
Subsequently, the information of each token is encoded in the form of a vector which can be learnt during training optimized by trainable parameters.
We consider the code tokens as a bag of words, because we need to avoid the influence of ordering among them.
For example, the embedding of ``read'' and ``file'' and that of ``file'' and ``read'' should be the same.
Thus, we use multiple fully connected layers as hidden layers to capture higher-level semantic information among the code tokens.
Then we sum all the vector representation of each token output by the last hidden layer as the final embedding of all the code tokens named token vector.\\

\noindent \emph{Joint Layer}
The joint layer takes as input the API context graph vector and the token vector and outputs a joint vector.
Suppose that the API context graph vector is a d$^A$-dimensional vector and the token vector is a d$^T$-dimensional vector.
The joint layer first combines the d$^A$-dimensional vector and the d$^T$-dimensional vector as a d$^{A+T}$-dimensional concat vector.
Then the d$^{A+T}$-dimensional concat vector is used to compute the final joint vector through a fully connected layer using \emph{tanh} as the activation function.
The fully connected layer is designed to further learn the joint semantics of the structural code information (in the form of the API context graph vector) and textual code information (in the form of the token vector) in a holistic way.
The joint vector output by the joint layer is used as the final vector for the softmax function. \\

\noindent \emph{Softmax Function}
In deep neural networks, the softmax function is usually used to map a vector to a normalized probability distribution over fixed size classes that needed to be predicted.
The classes are then ranked based on their probabilities.
If we consider each API as a class, the API recommendation task can be considered as a classification task.
What we need to do is to compute the probability of each API and then get the top $N$ APIs as the recommendations.
Thus, the softmax function is a natural choice. It takes as input the joint vector, and outputs a normalized probability over all APIs.

%% file: corpus.tex
\begin{algorithm}[t]
\caption{Training Instance Construction}\label{alg:trainSample}
\scriptsize
\begin{algorithmic}[1]
\REQUIRE $API\ context\ graph\ without\ a\ hole,\ node,\ hole\_size,\ code\_tokens$
\ENSURE $API\ context\ graph\ with\ a\ hole,\ remaining\ code\ tokens,\ label$
  \STATE $count=0$, $curr=node$
  \WHILE{$count$ is less than $hole\_size$ and $curr$ is not $Null$}
  	\STATE let $old$ be $curr$
	\IF {$curr$ is \textbf{If}, \textbf{While}, \textbf{Do}, \textbf{For}, \textbf{Foreach}, \textbf{Switch}, or \textbf{Try}}
		\FOR {each $child$ with edge type of $Type\ c$ or $Type\ cd$ of $curr$}
			\IF {$child$ represents the statement outside the control scope}
				\STATE $count=count+1$
				\STATE $curr=child$
			\ELSE
				\STATE remove $child$ and its subgraph
				\STATE remove all edges connected to $child$ and nodes in its subgraph
			\ENDIF
		\ENDFOR
	\ELSE	
		\STATE set $curr$ to be child with edge type of $Type\ c$ or $Type\ cd$ of $curr$
		\STATE $count=count+1$
	\ENDIF
	\STATE remove $old$ and all edges connected to $old$
  \ENDWHILE
  \STATE replace $node$ with \textbf{Hole}
  \STATE get remaining code tokens related to the $API\ context\ graph\ with\ a\ hole$
  \STATE set $label$ to be the label of $node$
\end{algorithmic}
\end{algorithm}
To train the models in~\app, we require a large set of training instances.
A training instance is a triple consisting of an API context graph (with a hole), a corresponding bag of code tokens and the expected label of the hole node (i.e., an API call).
To construct training instances, we first collect a large code base and then parse the methods one by one. For each method, we construct its corresponding API context graph (without a hole) and obtain the bag of code tokens.
Afterwards, we systematically replace a set of nodes from the API context graph with a hole node. The resultant API context graph (with a hole), the remaining code tokens and the label of the first removed node form a training instance.

The details of the algorithm for constructing a training instance is shown in Algorithm~\ref{alg:trainSample}. The inputs are an API context graph without a hole, the corresponding bag of code tokens, a node $node$ in the graph and a constant $hole\_size$. Intuitively, $node$ is the starting node to be removed and the label of $node$ is used as the label for the training instance, and $hole\_size$ is the number of nodes to be removed (including $node$) from the API context graph.

Algorithm~\ref{alg:trainSample} uses a variable $count$ to count the number of nodes that have been removed. Whenever $count$ reaches $hole\_size$ or there are no more nodes to be removed, the algorithm terminates.
Initially, we set $curr$ (which is the current node to be removed) to be $node$. If $curr$ is not a control node (like \emph{if} or \emph{while}), we identify its (unique) child node $child$ through an edge of \emph{Type c} or \emph{Type cd}, remove the current node $curr$ from the graph and set $curr$ to be $child$. Note that whenever a node is removed, so are its incoming and outgoing edges.
The reason why we choose the child node following edges of \emph{Type c} or \emph{Type cd} is that we remove nodes according to the control flow in the source code. As a result, the remaining context graph is still well-formed from a control flow point of view.
If $curr$ is a control node, all of its subgraphs in its control scope are removed, i.e., we remove all its subsequent nodes through control flow representing a statement in the control scope (e.g., all statements in the loop body if $curr$ is a \emph{while} node). For instance, if we remove the control node labeled with $While$ in Fig.~\ref{fig:get hash code from file}, all nodes representing the API call at line 6/7/8/9 are also removed, which are the ones labeled with $Condition,$ $java.io.BufferedReader.readLine(),$ $Body,$ 
$int.Declaration,$ $java.lang.String.hashCode()$ and $java.util.ArrayList.add(java.lang.Object)$.
Then, we set $curr$ to be the first subsequent node outside of the control scope.

For example, Fig.~\ref{fig:API context graph} is an API context graph with a hole that is produced from the code in Fig.~\ref{fig:get hash code from file}. 
In this example, the input of node $node$ is the node with label $java.lang.String.hashCode()$ representing the statement of $hashCode=str.hashCode();$ at line 8.
The input of hole size $hole\_size$ is set to be 1.
The input of code tokens $code\_tokens$ are all the tokens extracted in the original complete code.
The remaining code tokens are those tokens in the remaining source code, which are ``compute'', ``hash'', ``code'', ``path'', ``result'', ``rd'', ``br'' and ``str''.
For another instance, if all but line 2 and 3 are removed in Fig.~\ref{fig:get hash code from file}, the remaining code tokens become ``compute'', ``hash'', ``code'', ``path'', ``result'', and ``rd''.
The label of this training instance is $java.lang.String.hashCode()$.

To systematically construct a set of training instances, for each API context graph and code tokens constructed from a method in the code base, the above algorithm is applied with each node in the graph as the starting node to be removed and different hole sizes.
Note that the hole size can range from 1 to $Max - 1$ where $Max$ is the total number of nodes in the API context graph.

%% file: evaluation.tex
The purpose of \app~is recommending APIs based on given code context by combining structural and textual code information.
We develop an implementation of \app~for JDK 1.8, which has 17,173 API classes and 137,134 API methods/fields.
The implementation uses JavaParser~\cite{javaparser} to parse source code into ASTs and Java reflection mechanism to recognize API invocations in source code.
The lemmatization of code tokens is implemented using Stanford CoreNLP~\cite{Stanford_CoreNLP}.
The deep learning architecture is implemented using TensorFlow 1.14~\cite{tensorflow} and GG-NNs reference implementation~\cite{source_code_of_ggnn}.
Based on the implementation, we conduct a series of experimental studies to answer the following research questions.

\textbf{RQ1 (API Prediction Accuracy)}: How accurate is \app~in predicting the next API compared with state-of-the-art approaches for context-based API recommendation?

\textbf{RQ2 (Contribution of Textual Code Information)}: How much does textual code information contribute to the API recommendation?

\textbf{RQ3 (Effectiveness in Real Tasks)}: How effective is \app~in helping developers accomplish programming tasks?

All the data of the experimental studies can be found in our replication package~\cite{package}.

\subsection{Training Details}
\input{training}

\subsection{API Prediction Accuracy (RQ1)}
\input{rq1}

\subsection{Contribution of Textual Code Information (RQ2)}
\input{rq2}

\subsection{Effectiveness in Real Tasks (RQ3)} 
\input{rq3}

\subsection{Qualitative Analysis}\label{sec:qualitative analysis}
\input{qualitative_analysis}

\subsection{Threats to Validity}
\input{threats}

%% file: training.tex
We create a large corpus from GitHub by crawling all the Java projects that have 1000 stars or more.
In this way we obtain 1,914 Java projects, which include 944,783 source files, 7,279,321 methods, and 68,319,916 lines of code.

We randomly select 90\% of the Java projects as training set and the remaining 10\% projects as validation set.
For methods in the files of each project in the training or validation set, we apply Algorithm~\ref{alg:trainSample} to create a set of training instances or validation instances.
To ensure efficiency we filter out the files that are larger than 200 KB and the methods that have no JDK API invocations.
The reason for filtering files that are larger than 200 KB is that parsing large files using JavaParser~\cite{javaparser} is quite time consuming. Note that most of files (i.e, 99.9993\% of them) have a size smaller than 200KB and we expect filtering those large files has minimum effect. The reason for filtering methods that have no JDK API invocations is that we focus on JDK library.
When creating training/validation instances containing only preceding context, we do not limit the hole size (i.e., $hole\_size$);
when creating training/validation instances containing both preceding and succeeding contexts, we limit the hole size to 5 or less to avoid data explosion.
We also filter out training/validation instances that have no API invocation in the context.
Finally we obtain 6,627,591 training instances and 482,186 validation instances.

Based on the training data and validation data, we train an API recommendation model using a server with Intel Xeon E5-2620 2.1GHz (16 threads and 128GB RAM) and two Nvidia 1080Ti GPUs running on Ubuntu 16.04.
We set \textbf{embedding size} of each embedding layer to 300, the number of hidden layers to 3, \textbf{hidden size} of each hidden layer to 300, \textbf{dropout} to 0.75, \textbf{learning rate} to 0.005, and \textbf{batch size} to 256.
We conduct serval trial experiments with different hyper parameters and the above hyper parameters achieve the best performance.
After each epoch in the training, \app~evaluates the current model using the validation instances. 
If the prediction accuracy does not increase in five successive epochs, the training process ends and the last best model is used as the result.


%% file: rq1.tex
We compare \app~with existing approaches for solving the same problem.
We adopt two approaches that are most related to ours as baseline approaches in this evaluation. One is \GraLan~\cite{ICSE15GraLan}, which is a state-of-the-art graph-based statistical model for API recommendation and the other is \TreeLSTM~\cite{DeepAPIRec}, which is a state-of-the-art deep learning model using tree-based structure for API recommendation.
We reimplement \GraLan~based on the description of the approach in~\cite{ICSE15GraLan} and the extraction of graph representation from code in~\cite{Groum}.
The implementation of \TreeLSTM~is directly obtained from the authors of~\cite{DeepAPIRec}.
\app, \GraLan~ and \TreeLSTM~are trained with the same training data.
We choose six open-source Java projects as the test data: Galaxy~\cite{Galaxy}, Log4j~\cite{Log4j}, JGit~\cite{JGit}, Froyo-Email~\cite{Froyo-Email}, Grid-Sphere~\cite{Grid-Sphere}, and Itext~\cite{Itext}.
These projects are chosen based on the following criteria:
widely used as test data in previous researches on API recommendation (e.g.,~\cite{FSE16APIREC, ASE18RecRank});
not included in the training data or validation data.
Following the same procedure of training/validation instance construction, we create 14,986 test instances from the test data.

To confirm the effect of our \GraLan~implementation, we compare the API recommendation accuracy of our implementation on the six projects with that of the \GraLan~implementation by Liu et al.~\cite{ASE18RecRank} based on the results they report in~\cite{ASE18RecRank}.
The comparison shows that:
our implementation achieves a top-1 (top-10) accuracy of 19.6-41.6\% (73.4-80.9\%), while their implementation achieves a top-1 (top-10) accuracy of 22.4-33.6\% (73.9-80.6\%);
in terms of top-1 accuracy, our implementation is better than theirs on 4 projects and worse than theirs on 2 projects;
in terms of top-10 accuracy, our implementation is better than theirs on 4 projects and worse than theirs on 2 projects.
The results show that the performance of these two implementations is comparable.
Note that the performance of \GraLan~is sensitive to the count of each subgraph appeared in the training data.
Our training data is different from the training data used in~\cite{ASE18RecRank}, which explains why the performance of our implementation of \GraLan~is different from their original.


We compare the top-K accuracies and MRR (Mean Reciprocal Rank) of \app, \GraLan~and \TreeLSTM~for predicting the next API.
MRR is a summary metric for top-K accuracies that averages the inverse of the ranks of each recommendation, which ranges from 0 to 1~\cite{MRR}. For example, a MRR of 0.25 means that the correct recommendation is to appear at the fourth position on average.
The results are shown in Table~\ref{table:api prediction accuracy}.
In the table, the number of test instances of each project is shown after the project name and the best accuracy and MRR values are in boldface.
We can see that \app~achieves much higher top-1, top-5, and top-10 accuracy than \GraLan~and \TreeLSTM.
For the six projects, \app's top-1, top-5, and top-10 accuracy is 50.6-66.4\% (58.6\% on average), 67.7-87.1\% (81.4\% on average), and 79.2-92.5\% (87.9\% on average), respectively;
\GraLan's top-1, top-5, and top-10 accuracy is 19.6-41.6\% (31.5\% on average), 60.5-71.4\% (64.5\% on average), and 73.4-80.9\% (77.6\% on average), respectively;
\TreeLSTM's top-1, top-5, and top-10 accuracy is 39.3-52.6\% (46.7\% on average), 62.9-75.6\% (70.4\% on average), and 75.6-82.6\% (79.3\% on average), respectively.
We can see that \app~also achieves much higher MRR than \GraLan~and \TreeLSTM.
\app's MRR is 58.4-74.2\% (68.4\% on average), \GraLan's MRR is 37.4-53.8\% (45.3\% on average) and \TreeLSTM's MRR is 51.4-61.7\% (56.7\% on average). 
Furthermore, we conduct Mann-Whitney U test to determine whether the improvements in top-1, top-5, top-10 accuracy and MRR between \app~and the other two approaches are statistically significant.
If the p-value is less than 0.05, the improvement is considered to be significant.
The p-value of top-1, top-5 and top-10 accuracy between \app~and \GraLan~are 0.003, 0.004 and 0.004 respectively.
The p-value of top-1, top-5 and top-10 accuracy between \app~and \TreeLSTM~are 0.015, 0.023 and 0.010 respectively.
The p-value of MRR between \app~and \GraLan~is 0.003 and the p-value of MRR between \app~and \TreeLSTM~is 0.007.
We can see that all the improvements are significant.

\begin{table}
  \scriptsize
  \centering
  \caption{API Recommendation's Top-K Accuracy and MRR (\%)}
  \label{table:api prediction accuracy}
  \begin{tabular}{|c|c|c|c|c|c|c|c|}
    \hline
    \textbf{Project} &
    \textbf{Model} &
    \textbf{Top-1} &
    \textbf{Top-5} &
    \textbf{Top-10} &
    \textbf{MRR}\\
    \hline
    Galaxy & \GraLan & 29.4 & 60.5 & 73.4 & 42.2\\
    (473)  & \TreeLSTM & 39.3 & 68.7 & 76.7 & 51.4\\
      & \app & \textbf{51.0} & \textbf{81.6} & \textbf{88.2} & \textbf{63.6}\\
    \hline
    JGit & \GraLan & 41.6 & 71.4 & 79.1 & 53.8\\
    (4530) & \TreeLSTM & 52.6 & 75.6 & 81.2 & 61.7\\
    & \app & \textbf{66.4} & \textbf{85.1} & \textbf{89.5} & \textbf{74.2}\\
    \hline
    Froyo-Email  & \GraLan & 23.0 & 62.8 & 78.9 & 40.7\\
    (1537)  & \TreeLSTM & 51.6 & 74.5 & 82.6 & 61.1\\
    & \app & \textbf{63.7} & \textbf{86.0} & \textbf{91.3} & \textbf{73.5}\\ 
    \hline
    Grid-Sphere & \GraLan & 36.5 & 66.5 & 80.9 & 48.6\\
    (1847) & \TreeLSTM & 48.0 & 72.4 & 80.6 & 58.3\\
    & \app & \textbf{62.0} & \textbf{87.1} & \textbf{92.5} & \textbf{72.8}\\
    \hline
    Itext  & \GraLan & 19.6 & 64.3 & 75.7 & 37.4\\
    (4444) & \TreeLSTM & 46.0 & 68.1 & 75.6 & 55.4\\
    & \app & \textbf{57.9} & \textbf{80.7} & \textbf{86.8} & \textbf{67.6}\\ 
    \hline
    Log4j & \GraLan & 38.6 & 61.3 & 77.5 & 48.9\\
    (2155) & \TreeLSTM & 42.4 & 62.9 & 79.0 & 52.2\\
     & \app & \textbf{50.6} & \textbf{67.7} & \textbf{79.2} & \textbf{58.4}\\
    \hline
    \multirow{3}{*}{Average}  & \GraLan & 31.5 & 64.5 & 77.6 & 45.3\\
     & \TreeLSTM & 46.7 & 70.4 & 79.3 & 56.7\\
     & \app & \textbf{58.6} & \textbf{81.4} & \textbf{87.9} & \textbf{68.4} \\ 
    \hline
  \end{tabular}
\end{table}

%% file: rq2.tex
\begin{table*}[t]
  \scriptsize
  \centering
  \caption{Contribution of Textual Code Information(\%)}
  \label{table:contribution of textual code information}
  \begin{tabular}{|c|c|c|c|c|c|c|c|c|c|c|}
    \hline
    \textbf{Project} &
    \textbf{Model} &
    \textbf{Top-1} &
    \textbf{Difference} &
    \textbf{Top-5} &
    \textbf{Difference} &
    \textbf{Top-10} &
    \textbf{Difference} &
    \textbf{MRR} &
    \textbf{Difference} \\
    \hline
    Galaxy & \singleModel & 46.9 & \multirow{2}{*}{+4.1} & 76.3 & \multirow{2}{*}{+5.3} & 82.2 & \multirow{2}{*}{+6.0} & 58.9 & \multirow{2}{*}{+4.7}\\
    (473)  & \app & \textbf{51.0} & & \textbf{81.6} & & \textbf{88.2} & & \textbf{63.6} & \\
    \hline
	JGit & \singleModel & 61.7 & \multirow{2}{*}{+4.7} & 83.8 & \multirow{2}{*}{+1.3} & 88.6 & \multirow{2}{*}{+0.9} & 71.2 & \multirow{2}{*}{+3.0}\\
    (4530)  & \app & \textbf{66.4} & & \textbf{85.1} & & \textbf{89.5} & & \textbf{74.2} & \\
    \hline
	Froyo-Email & \singleModel & 58.8 & \multirow{2}{*}{+4.9} & 82.4 & \multirow{2}{*}{+3.6} & 88.9 & \multirow{2}{*}{+2.4} & 68.7 & \multirow{2}{*}{+4.8} \\
    (1537)  & \app & \textbf{63.7} & & \textbf{86.0} & & \textbf{91.3} & & \textbf{73.5} & \\
    \hline
	Grid-Sphere & \singleModel & 57.7 & \multirow{2}{*}{+4.3} & 83.1 & \multirow{2}{*}{+4.0} & 90.6 & \multirow{2}{*}{+1.9} & 69.0 & \multirow{2}{*}{+3.8}\\
    (1847)  & \app & \textbf{62.0} & & \textbf{87.1} & & \textbf{92.5} & & \textbf{72.8} & \\
    \hline
	Itext & \singleModel & 56.3 & \multirow{2}{*}{+1.6} & 78.8 & \multirow{2}{*}{+1.9} & 84.4 & \multirow{2}{*}{+2.4} & 65.7 & \multirow{2}{*}{+1.9}\\
    (4444)  & \app & \textbf{57.9} & & \textbf{80.7} & & \textbf{86.8} & & \textbf{67.6} & \\
    \hline
     Log4j & \singleModel & 48.5 & \multirow{2}{*}{+2.1} & \textbf{71.4} & \multirow{2}{*}{-3.7} & \textbf{83.7} & \multirow{2}{*}{-4.5} & 58.3 & \multirow{2}{*}{+0.1}\\
    (2155)  & \app & \textbf{50.6} & & 67.7 & & 79.2 & & \textbf{58.4} & \\
    \hline
    Overall & \singleModel & 56.9 & \multirow{2}{*}{+3.4} & 80.0 & \multirow{2}{*}{+1.5} & 86.7 & \multirow{2}{*}{+1.0} & 66.8 & \multirow{2}{*}{+2.6}\\
    (14986)  & \app & \textbf{60.3} & & \textbf{81.5} & & \textbf{87.7} & & \textbf{69.4} & \\
    \hline
    \end{tabular}
\end{table*}

\app~mainly relies on the structural code information embedded in the \APINetwork~and at the same time leverages the textual code information embedded in the \TokenNetwork.
To evaluate the contribution of textual code information, we derive a variant of \app~that uses structural code information only (called \singleModel), which only includes one network (i.e., \APINetwork).
We use \singleModel~to train an API recommendation model based on the same training/validation data and evaluate the model with the same test data.
The results are shown in Table~\ref{table:contribution of textual code information}.
We can see  that \singleModel~achieves good top-1, top-5, and top-10 overall accuracy (56.9\%, 80.0\%, and 86.7\%) on the six projects, but the accuracy is lower than that of \app~(60.3\%, 81.5\%, and 87.7\%).
The top-1 overall accuracy achieves a 3.4\% improvement, the top-5 overall accuracy achieves a 1.5\% improvement and the top-10 overall accuracy achieves a 1.0\% improvement when textual code information is added.
For each test project, the top-k accuracy of adding textual code information achieves different degrees of improvement.
The improvement of the top-1 accuracy ranges from 1.6\% to 4.9\%, the improvement of the top-5 accuracy ranges from 1.3\% to 5.3\%, and the improvement of the top-10 accuracy ranges from 0.9\% to 6.0\%. 
We can also see that the overall MRR is improved by 2.6\% when textual code information is added.
For each test project, the improvement of MRR ranges from 0.1\% to 4.8\%.
The top-5 and top-10 accuracy of Log4j project decrease when textual code information is added.
It is because that textual code information maybe contains noise that negatively influences the API recommendation results.
In our future work, we will try to better process the noise in textual code information.

\begin{figure*}
    \centering
    \includegraphics[scale=0.55]{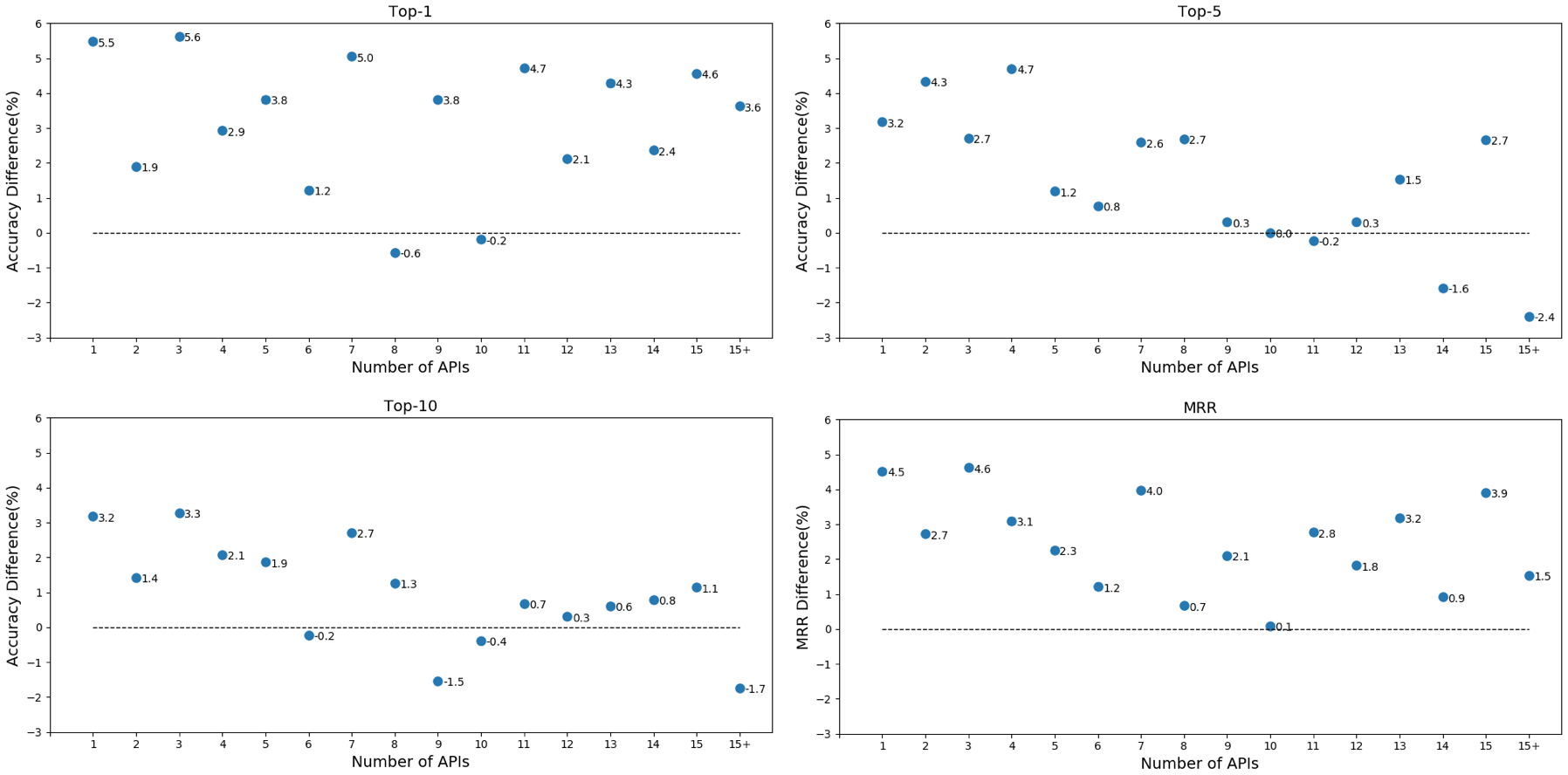} 
    \caption{Textual Code Information Contribution with the Increase of API Numbers in the Context}
    \label{fig:textContribution} 
\end{figure*}

To further understand the contribution of textual code information, we analyze its influence with an increasing number of APIs in the context.
We divide all the test data into 16 subsets according to the number of APIs in the context (1-15 and above 15).
For each subset, we calculate the difference of the top-1, top-5, top-10 accuracy and MRR of \app~and~\singleModel.
The results are shown in Figure~\ref{fig:textContribution}.
The dotted lines are the zero lines and the points above the lines indicate positive contribution of textual code information, which mean that \app~achieves higher accuracy and MRR than~\singleModel.
We can see that the contribution of textual code information is positive in most cases.
There is no obvious positive or negative correlation between the contribution of textual code information and the number of APIs in the context.
This means that the contribution of textual code information is insensitive to the number of APIs in the context.
The reason of the nine negative cases in Figure~\ref{fig:textContribution} is also that textual code information maybe contains noise that negatively influences the API recommendation results.

%% file: rq3.tex
We develop an IntelliJ IDEA plugin for \app~and conduct a user study in which two groups of participants are asked to complete a set of programming tasks with and without the plugin respectively.
Note that the purpose of the user study is not to compare \app~with other approaches, since we have already answered RQ1. The objective is rather to evaluate whether \app~can indeed help developers during coding. So, two groups of participants are asked to complete a set of programming tasks with and without using the \app's~plugin respectively.
We derive a set of programming tasks from Stack Overflow (SO) in the following way.
We find the 500 most voted SO questions with the tag ``Java'' and identify those that can be used as programming tasks.
For example, questions about concept explanation such as ``Is Java pass-by-reference or pass-by-value?'' are eliminated.
We then choose those questions that have code snippets in the answers or question bodies that can be used to implement the desired functionalities.
We further filter out the questions that have less than four lines of code or are not API intensive.
We obtain 44 SO questions as candidates and randomly select the following six as the tasks.
For each task we prepare a description based on the corresponding question title and body and design a set of test cases (2-9, 6 on average).

T1: How do I create a Java string from the contents of a file~\cite{t1}

T2: Iterating through a Collection, avoiding ConcurrentModificationException when removing objects in a loop~\cite{t2}

T3: How can I generate an MD5 hash~\cite{t3}
 
T4: How do I invoke a Java method when given the method name as a string~\cite{t4}

T5: How to read all files in a folder from Java~\cite{t5}

T6: How can I increment a date by one day in Java~\cite{t6}

We recruit 18 master students from our school and all of them major in software engineering.
Based on a pre-experiment survey on their experience with Java programming, we divide them into two groups whose overall abilities are at an equivalent level.
We respectively assign G1 to use standard IntelliJ IDEA and G2 to use IntelliJ IDEA with the \app~plugin.
The participants are asked to complete the six tasks from T1 to T6.
They are not allowed to search Internet, but can look up the JDK reference documentation and use the code recommendation feature and other facilities provided by IntelliJ IDEA.
The participants in G2 can request the help of the \app~plugin, which can provide a list of top 10 API recommendations for the current cursor position.
For each task the participants are given 20 minutes and if they cannot finish it in time they have to stop and submit their implementation.
We record the completion time of the participants and test their implementations for each task.

We use task completion time and test pass rate as two metrics for evaluation. Task completion time is the time that a participant used to complete a task. Given a submitted implementation of a task, test pass rate is the percentage of test cases passed in the total number of test cases.
The results of descriptive statistics analysis of task completion time and test pass rate are shown in Table~\ref{tb:time} and Table~\ref{tb:pass} respectively.
On average, the participants in G1 use 665.7-1,173.3 seconds to finish a task, while the participants in G2 use 441.3-708.1 seconds to finish a task;
the participants in G1 pass 4-47\% test cases, while the participants in G2 pass 68-89\% test cases.
We can see that \app~helps the participants finish the tasks faster and more accurately.
Furthermore, we evaluate whether the improvements are significant or not.
We make a significance test using the Mann-Whitney U test where a difference is thought to be significant if the p-value is less than 0.05.
We can see that the participants in G2 significantly outperform the participants in G1 in terms of completion time for three tasks and in terms of test pass rate for five tasks.

We have an interview with each of the participants in G2 to get their feedback on \app.
Most of them agree that \app~provides accurate recommendations which are quite helpful especially when they do not know how to proceed. 
In most cases, the right API is included in the top 5 recommendations.
In extreme cases, \app~can even provide right APIs when the participants only declare a method (including method name and parameters).
This indicates that \app~can provide useful recommendations by only using textual code information.
They also provide suggestions for further improvement.
Two common suggestions are recommending arguments for API invocation and providing explanations for the recommended APIs.

\begin{table}[t]
  \scriptsize
  \centering
  \caption{Completion Time of the Tasks (Seconds)}
  \label{tb:time}
  \begin{tabular}{|c|c|c|c|c|c|c|c|}
    \hline
    Task &
    Group &
    avg & 
    min & 
    max &
    median &
    stan. dev.&
    p \\
    \hline
    \multirow{2}{*}{T1} & 
    G1 & 888.0& 485& 1200& 900& 299.83 & \multirow{2}{*}{0.0904}  \\
     \cline{2-7}
    & G2 & 680.0 & 188 & 1200 & 718 & 343.13  & \\
    \hline
    \multirow{2}{*}{T2} & 
    G1 & 671.4 & 232 & 1200 & 480 & 387.91 & \multirow{2}{*}{0.2677}  \\
     \cline{2-7}
    & G2 & 562.4 & 246 & 1003 & 449 & 298.53 & \\
    \hline
    \multirow{2}{*}{T3} & 
    G1 & 1173.3 & 960 & 1200 & 1200 & 75.42 & \multirow{2}{*}{0.0001}  \\
     \cline{2-7}
    & G2 & 441.3 & 160 & 703 & 463 & 150.04 & \\
    \hline
    \multirow{2}{*}{T4} & 
    G1 & 1159.6 & 836 & 1200 & 1200 & 114.39 & \multirow{2}{*}{0.0003}  \\
     \cline{2-7}
    & G2 & 708.1 & 431 & 1154 & 697 & 211.87 & \\
    \hline
    \multirow{2}{*}{T5} & 
    G1 & 665.7 & 345 & 1200 & 558 & 295.51 & \multirow{2}{*}{0.0924}  \\
     \cline{2-7}
    & G2 & 475.0 & 232 & 746 & 427 & 184.56 & \\
    \hline
    \multirow{2}{*}{T6} & 
    G1 & 1140.0& 660 & 1200 & 1200 & 169.71 & \multirow{2}{*}{0.0013}  \\
     \cline{2-7}
    & G2 & 707.3 & 255 & 1200 & 658 & 340.67  & \\
    \hline
    \end{tabular}
\end{table}

\begin{table}[t]
  \scriptsize
  \centering
  \caption{Test Pass Rate of the Tasks}
  \label{tb:pass}
  \begin{tabular}{|c|c|c|c|c|c|c|c|}
    \hline
    Task &
    Group &
    avg & 
    min & 
    max &
    median &
    stan. dev.&
    p-value \\
    \hline
    \multirow{2}{*}{T1} & 
    G1 & 0.26 & 0.00 & 1.00 & 0.00 & 0.41 & \multirow{2}{*}{0.0073}  \\
     \cline{2-7}
    & G2 & 0.81 & 0.00 & 1.00 & 1.00 & 0.36  & \\
    \hline
    \multirow{2}{*}{T2} & 
    G1 & 0.47 & 0.00 & 1.00 & 0.33 & 0.41 & \multirow{2}{*}{0.1461}  \\
     \cline{2-7}
    & G2 & 0.68 & 0.00 & 1.00 & 1.00 & 0.39 & \\
    \hline
    \multirow{2}{*}{T3} & 
    G1 & 0.11 & 0.00 & 1.00 & 0.00 & 0.31 & \multirow{2}{*}{0.0008}  \\
     \cline{2-7}
    & G2 & 0.89 & 0.00 & 1.00 & 1.00 & 0.31 & \\
    \hline
    \multirow{2}{*}{T4} & 
    G1 & 0.11 & 0.00 & 1.00 & 0.00 & 0.31 & \multirow{2}{*}{0.0012}  \\
     \cline{2-7}
    & G2 & 0.83 & 0.00 & 1.00 & 1.00 & 0.33 & \\
    \hline
    \multirow{2}{*}{T5} & 
    G1 & 0.36 & 0.00 & 1.00 & 0.38 & 0.37 & \multirow{2}{*}{0.0030}  \\
     \cline{2-7}
    & G2 & 0.89 & 0.5 & 1.00 & 1.00 & 0.21 & \\
    \hline
    \multirow{2}{*}{T6} & 
    G1 & 0.04 & 0.00 & 0.33 & 0.00 & 0.10 & \multirow{2}{*}{0.0013}  \\
     \cline{2-7}
    & G2 & 0.78 & 0.00 & 1.00 & 1.00 & 0.42  & \\
    \hline
    \end{tabular}
\end{table}

%% file: qualitative_analysis.tex
\begin{figure*}
    \centering
    \includegraphics[scale=0.50]{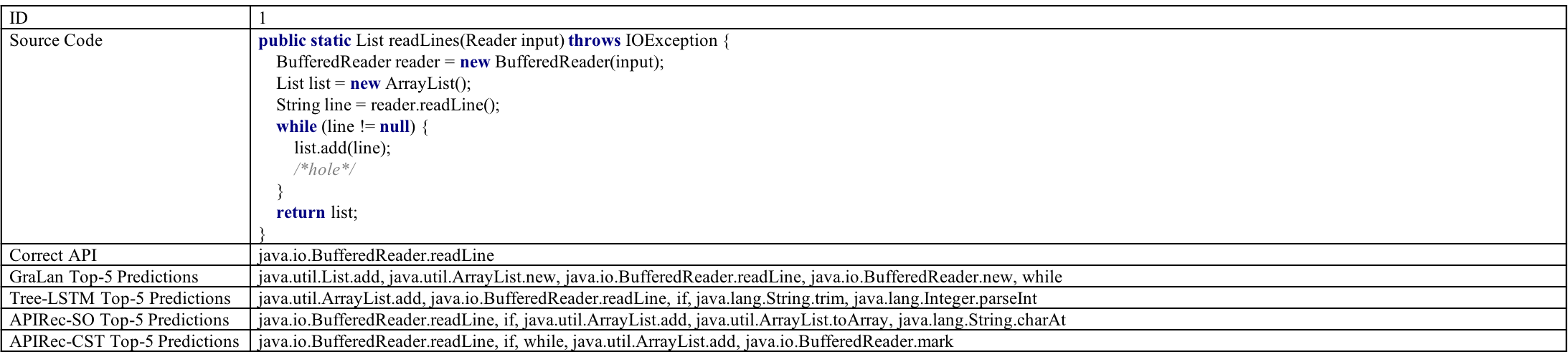} 
    \includegraphics[scale=0.50]{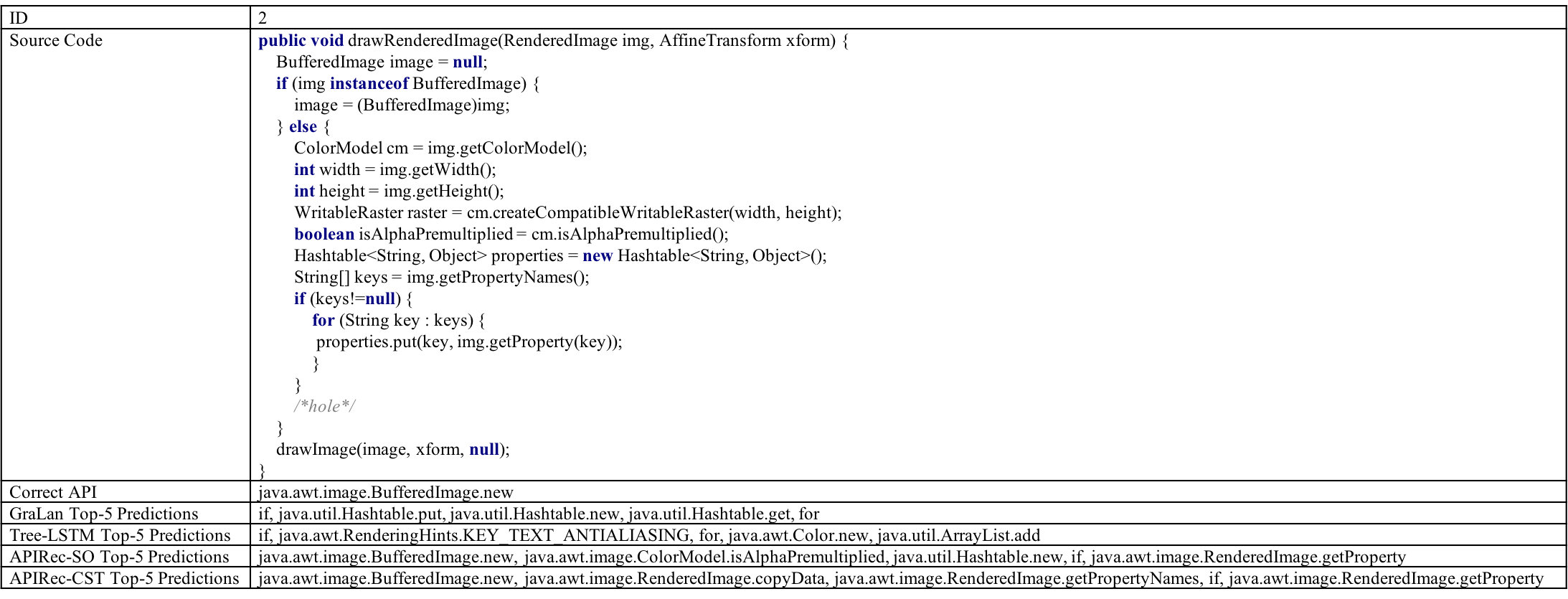}
    \includegraphics[scale=0.50]{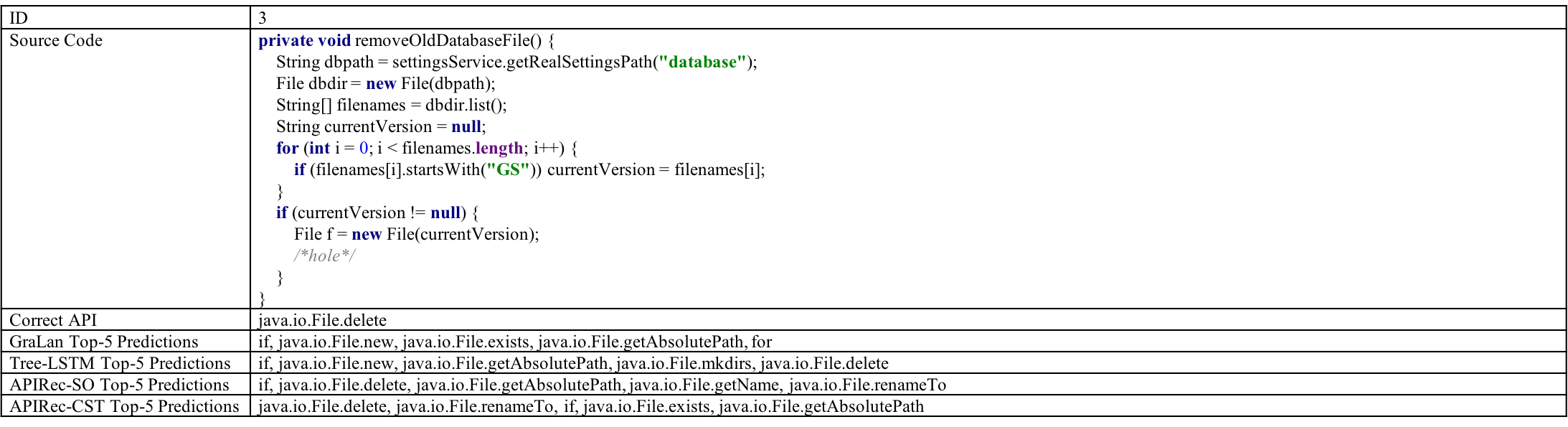}
    \includegraphics[scale=0.50]{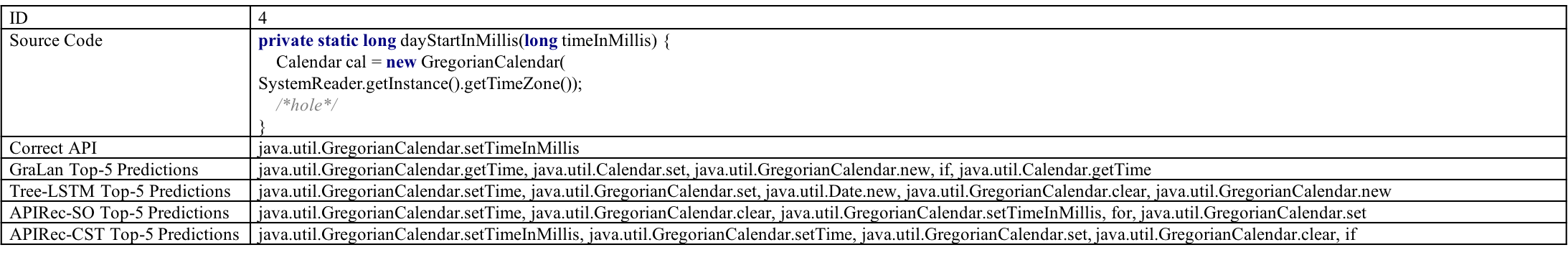}
    \includegraphics[scale=0.50]{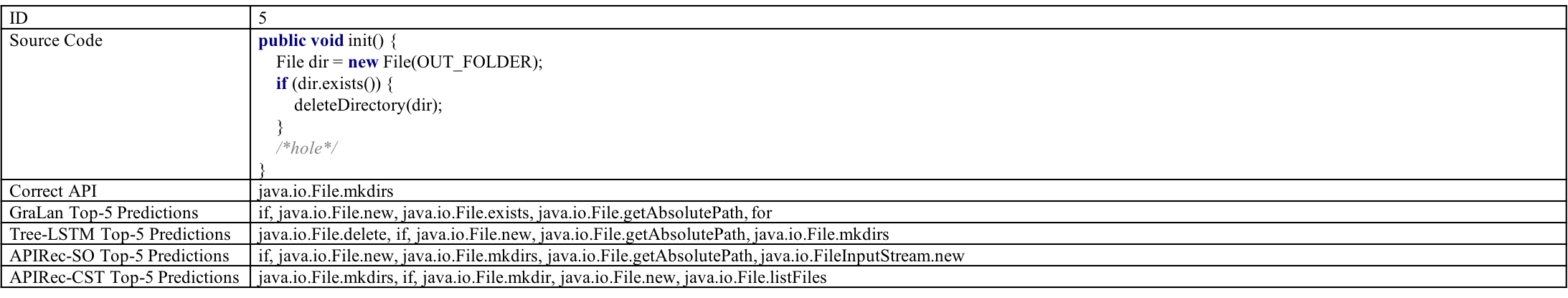}   
    \caption{Qualitative Analysis Examples}
    \label{fig:qualitative analysis examples} 
\end{figure*}

In RQ1 and RQ2, we perform quantitative analysis on \app, thus we list some examples to qualitatively illustrate the advantages of \app.
In Figure~\ref{fig:qualitative analysis examples}, we list five examples.

The first example is to read contents from a reader of a file line by line.
As we can see that \GraLan~recommends the correct API in the third place, \TreeLSTM~recommends the correct API in the second place, and both \singleModel~and \app~recommend the correct API in the first place.
The first two recommendations of \GraLan~are due to the irrelevant subgraphs that capture the semantics of list operation.
This suggests that though subgraphs in \GraLan~may capture the semantics at a hole, the recommendations may be over-shadowed by other irrelevant subgraphs.
\TreeLSTM~is a deep learning model using tree-based structure which includes control flow among APIs but lack of data flow.
\TreeLSTM~treats source code as code tree, and feed the code tree into the deep learning model.
Compared to \GraLan, \TreeLSTM~also considers the structure information but lack of data flow.
Thus, \TreeLSTM~performs better than \GraLan, although worse than \singleModel~and \app.
\singleModel~and \app~treat source code as an API context graph which contains the structure information, and apply GG-NNs to learn the semantic in an API context graph using a holistic view.
Due to the information diffusion mechanism in GG-NNs, each node itself and its relations of other nodes in the API context graph are integrated and  added to the final vector representation of the API context graph.
As a result, \singleModel~and \app~successfully recommend that the API of the hole should be used to read the next line.
From this example, we can see that a holistic view of correlated API usage in control and data flow graph of an entire method can help to improve the ranking of the correct API.

The second example is to draw a BufferedImage given a RenderedImage.
As we can see that \GraLan~and \TreeLSTM~fail to recommend the correct API in the top 5 recommendations, whereas \singleModel~and \app~successfully recommend the correct API in the first place.
In this example, none of the subgraphs in \GraLan~can capture the real semantics at the hole.
Most of \GraLan's recommendations are the APIs in $java.util.Hashtable$ because APIs in $java.util.Hashtable$ are closest to the hole and are used as context in subgraphs.
Due to the lack of data flow among APIs, \TreeLSTM~cannot recommend the correct API.
Only with a holistic view of a control and data flow graph of an entire method, can \singleModel~and \app~find that all the APIs in the method are prepared to be used as the parameters of the correct API $java.awt.image.BufferedImage.new(java.awt.image.C$
$olorModel,java.awt.image.WritableRaster,boolean,jav$
$a.util.Hashtable)$ to create a BufferedImage object.
From this example, we can see that in some situations, a holistic view of correlated API usage in control and data flow graph of an entire method
can help to recommend the correct API.

The third example is to remove an old database file.
As we can see that \GraLan~fails to recommend the correct API in the top 5 recommendations, \TreeLSTM~recommends the correct API in the fifth place, whereas \singleModel~recommends the correct API in the second and \app~recommends the correct API in the first place.
All of the approaches capture the semantics at the hole is to apply an operation to a File object. 
However, the first three approaches fail to identify which operation should be applied to the File object.
\app~leverages the method name as textual information in which ``remove'' indicates that the operation is to delete a file.
\app~applies a \TokenNetwork~to embed the textual information to capture the semantics in the textual information and combined (joint) with the structure information.
From this example, we can see that the method name is indeed helpful to clarify the semantics.

The fourth example is to set the time in millisecond of a given value.
As we can see that \GraLan~and \TreeLSTM~fail to recommend the correct API in the top 5 recommendations, \singleModel~recommends the correct API in the third place and \app~recommends the correct API in the first place.
Since there is only one JDK API (Calendar) in the method, recommendations of all the approaches are related to the Calendar object. 
However, the first three approaches cannot recommend the correct API in first place because they are not certain which operation should be applied on the Calendar object.
\app~leverages the parameter name as textual information in which ``time'', ``in'' and ``millis'' (``in'' and ``millis'' are also in the method name) indicate that the operation is to process time in millisecond.
Combined with the semantics in the API context graph, \app~successfully identifies that the operation is to set the time in millisecond.
From this example, we can see that the parameter name is helpful to clarify the semantics.

The last example is to create a new directory (delete the original directory if the directory exists).
As we can see that \GraLan~fail to recommend the correct API in the top 5 recommendations, \TreeLSTM~recommends the correct API in the fifth place, \singleModel~recommends the correct API in the third place and \app~recommends the correct API in first place.
All of the approaches capture the semantics at the hole is to apply an operation to a File object. 
However, the first three approaches fail to identify which operation should be applied to the File object, and thus fail to recommend the correct API in the first place.
\app~leverages the variable names as textual information in which ``dir'' and ``folder'' indicate that the operation should applied on a directory not a file.
Combined with the semantics in the API context graph, \app~successfully identifies that there lacks a new directory and thus the operation is to create a new directory.
From this example, we can see that the variable names are helpful to clarify the semantics.

%% file: threats.tex
The threats to the internal validity of our studies lie in two aspects.
First, the \GraLan~implementation may not be exactly consistent with the approach.
Second, the test cases developed for each task may not be complete.

The threats to the external validity of our studies lie in two aspects.
First, we only implement our approach for Java and evaluate it with JDK.
It is not clear how well the approach can support other languages and API libraries. 
Second, as adopted in~\cite{ICSE15GraLan,DeepAPIRec,ASE18RecRank}, the test cases used in RQ1 are constructed automatically which may not reflect the scenarios in the real world. Different from existing approaches, we additionally conduct a user study to simulate the scenarios in the real world to evaluate the effectiveness of \app.
Third, we only evaluate the approach with a group of master students and a set of tasks from SO in the user study.
It is not clear how effectively the approach can support industrial developers to accomplish more complex programming tasks.

%% file: related_work.tex
This work is closely related to various research on code recommendation. 
In modern IDE (Integrated Development Environment), type information is often used to recommend API method calls when classes or objects are typed.
To enhance the performance of the code recommendation in IDEs, several approaches have been proposed to sort, filter and group API methods for better recommendation~\cite{ICSM09David,ICSE10Daqing,ICSM11Daqing}.
In comparison, \app~does not require a developer to write a receiver expression.
Heinemann \emph{et al.}~\cite{identifier-based} propose an API method recommendation algorithm based on the extracted identifiers (such as variable and type names) in the development context.
In comparison, in addition to the textual information (including identifiers), \app~also takes structural information into consideration.
Besides, \app~uses a deep neural network to learn the semantics of the textual information instead of simply using Jaccard similarity.
Several approaches~\cite{BMN,CSCC} compute the similarity between the current code context and previous code examples based on a set of API calls or other additional information (such as method names, Java keywords, class or interface names).
In comparison, \app~considers the complete API usage modeled in an API context graph, which contains API calls, Java keywords, and control and data flow among them. Furthermore, \app~combines textual information which includes method names, parameter names, and variable names.

This work is also related to work on mining usage patterns from source code, such as~\cite{CoRR15APIMining,MSR13APIMing,MSR06APIMining,FSE09GrouMiner,ICSE12Grapacc}.
These approaches often apply deterministic mining algorithms to mine usage patterns for code recommendation.
Zhong \emph{et al.}~\cite{MSR06APIMining} propose MAPO to cluster code snippets and mine usage patterns by frequent subsequence mining.
Nguyen \emph{et al.}~\cite{FSE09GrouMiner} propose GrouMiner to mine usage patterns by representing source code as groums.
Nguyen \emph{et al.} propose Grapacc~\cite{ICSE12Grapacc}, which first mines usage patterns based on graphs and then matches these patterns with the code fragment under editing based on graph-based features and token-based features.
Wang \emph{et al.}~\cite{MSR13APIMing} apply a two-step clustering strategy to cluster call sequences and mine usage patterns for each cluster using a frequent closed sequence mining algorithm.
Fowkes \emph{et al.}~\cite{CoRR15APIMining} propose a near parameter-free probabilistic algorithm to infer the most interesting usage patterns.
In comparison, \app~learn regularity of the API usage based on deep learning techniques instead of mining usage patterns explicitly.

Based on the conjecture that source code is naturally repetitive and predictable~\cite{ICSE12Naturalness}, many approaches have been proposed to learn statistical language models from source code for code recommendation.
Hindle \emph{et al.}~\cite{ICSE12Naturalness} train an n-gram model based on the tokens of the source code to recommend the next token.
Allamanis \emph{et al.}~\cite{MSR13tatisticalLearninig} use a large corpus of source code from various domains to train an n-gram model.
Nguyen \emph{et al.}~\cite{FSE13SLAMC} enhance the n-gram model with roles and data types of code tokens and global technical concerns/functionality.
Tu \emph{et al.}~\cite{FSE14Localness} enhance the n-gram model with a cache to capture the localized regularities in the source code to improve the accuracy.
Nguyen \emph{et al.}~\cite{ICSE15GraLan} propose a graph-based statistical language model by using Bayesian statistical inference to compute the probabilities of API recommendations based on graphs.
Liu \emph{et al.}~\cite{ASE18RecRank} propose a re-ranking approach based on the top-10 recommendations of \GraLan~to improve the top-1 accuracy using API usage paths as features.
Nguyen \emph{et al.}~\cite{FSE16APIREC} propose APIRec that learns from fine-grained code changes by developing an association-based change inference model to recommend API calls.
In comparison, \app~learns from the control and data flow in the source code instead of treating the source code as tokens as in the above-mentioned proposals.
Furthermore, \app~takes a holistic approach to learn from both structural and textual code information.

There are also approaches which apply deep learning techniques for code recommendation. Raychev \emph{et al.}~\cite{PLDI14SLANG} treat the source code as sentences and combine the n-gram model with RNN for recommending sentences.
Dam \emph{et al.}~\cite{LSTMForCode} train an LSTM (Long Short-Term Memory) neural network based on code tokens.
Nguyen \emph{et al.}~\cite{SANER18Dnn4C} train a deep neural network called Dnn4C, which not only leverages the local context of lexical code elements, but also syntactic and type contexts.
In comparison, \app~combines a graph-based deep neural network and a token-based deep neural network to capture both structural and textual code information.

This work is broadly related to other applications of deep learning techniques on source code for various objectives including code summarization~\cite{DL4SourceCode_1}, code generation~\cite{CodeGeneration_1,CodeGeneration_2,CodeGeneration_3,CodeGeneration_4,CodeGeneration_5}, code search~\cite{DeepAPILearning,DeepCS}, comment generation~\cite{CommentGeneration_1,CommentGeneration_2,CommentGeneration_3,CommentGeneration_4,CommentGeneration_5} or defect prediction~\cite{DefectPrediction_1,DefectPrediction_2}.
For example, Allamanis \emph{et al.}~\cite{DL4SourceCode_1} propose an attentional neural network to give an extreme summary of a sequence of code tokens.
Mou \emph{et al.}~\cite{CodeGeneration_5} apply a sequence-to-sequence recurrent neural network to generate code when given a user intention. 
Gu \emph{et al.}~\cite{DeepCS} propose a deep neural network called CODEnn to jointly embed code snippets and natural language descriptions.
Hu \emph{et al.}~\cite{CommentGeneration_1} propose DeepCom which takes AST sequences of source code as input and generates the corresponding comments based on an attentional Seq2Seq model.
Wang \emph{et al.}~\cite{DefectPrediction_2} apply Deep Belief Network to learn features of tokens extracted from source code for defect prediction.
These approaches apply different deep learning models to learn program semantics for different objectives. In comparison, \app~represents program as an API context graph and a bag of code tokens and designs a novel deep neural network for API recommendation.

%% file: conclusion.tex
In this paper we propose a deep learning based API recommendation approach that combines the API usage with the text information in the source code to simultaneously learn structural and textual features. 
Our evaluation shows that our approach significantly outperforms an existing graph-based statistical model and a tree-based deep learning model for API recommendation and can effectively help students to finish programming tasks faster and more accurately.
Our future work will improve the approach from several aspects.
First, we will improve the utilization of textual code information, for example by using better data preprocessing methods and model architectures or introducing user interactions.
Second, we will incorporate argument recommendation and API explanation into the approach.
Third, we will apply our approach for other API libraries and try to extend the approach to support API recommendation of multiple libraries.

%% file: main.bbl
\begin{thebibliography}{10}
\providecommand{\url}[1]{#1}
\csname url@samestyle\endcsname
\providecommand{\newblock}{\relax}
\providecommand{\bibinfo}[2]{#2}
\providecommand{\BIBentrySTDinterwordspacing}{\spaceskip=0pt\relax}
\providecommand{\BIBentryALTinterwordstretchfactor}{4}
\providecommand{\BIBentryALTinterwordspacing}{\spaceskip=\fontdimen2\font plus
\BIBentryALTinterwordstretchfactor\fontdimen3\font minus
  \fontdimen4\font\relax}
\providecommand{\BIBforeignlanguage}[2]{{%
\expandafter\ifx\csname l@#1\endcsname\relax
\typeout{** WARNING: IEEEtran.bst: No hyphenation pattern has been}%
\typeout{** loaded for the language `#1'. Using the pattern for}%
\typeout{** the default language instead.}%
\else
\language=\csname l@#1\endcsname
\fi
#2}}
\providecommand{\BIBdecl}{\relax}
\BIBdecl

\bibitem{ICSE12Naturalness}
A.~Hindle, E.~T. Barr, Z.~Su, M.~Gabel, and P.~T. Devanbu, ``On the naturalness
  of software,'' in \emph{34th International Conference on Software
  Engineering, {ICSE} 2012, June 2-9, 2012, Zurich, Switzerland}, 2012, pp.
  837--847.

\bibitem{MSR13tatisticalLearninig}
M.~Allamanis and C.~A. Sutton, ``Mining source code repositories at massive
  scale using language modeling,'' in \emph{Proceedings of the 10th Working
  Conference on Mining Software Repositories, {MSR} '13, San Francisco, CA,
  USA, May 18-19, 2013}, 2013, pp. 207--216.

\bibitem{FSE13SLAMC}
T.~T. Nguyen, A.~T. Nguyen, H.~A. Nguyen, and T.~N. Nguyen, ``A statistical
  semantic language model for source code,'' in \emph{Joint Meeting of the
  European Software Engineering Conference and the {ACM} {SIGSOFT} Symposium on
  the Foundations of Software Engineering, ESEC/FSE'13, Saint Petersburg,
  Russian Federation, August 18-26, 2013}, 2013, pp. 532--542.

\bibitem{FSE14Localness}
Z.~Tu, Z.~Su, and P.~T. Devanbu, ``On the localness of software,'' in
  \emph{Proceedings of the 22nd {ACM} {SIGSOFT} International Symposium on
  Foundations of Software Engineering, (FSE-22), Hong Kong, China, November 16
  - 22, 2014}, 2014, pp. 269--280.

\bibitem{PLDI14SLANG}
V.~Raychev, M.~T. Vechev, and E.~Yahav, ``Code completion with statistical
  language models,'' in \emph{{ACM} {SIGPLAN} Conference on Programming
  Language Design and Implementation, {PLDI} '14, Edinburgh, United Kingdom -
  June 09 - 11, 2014}, 2014, pp. 419--428.

\bibitem{LSTMForCode}
H.~K. Dam, T.~Tran, and T.~Pham, ``A deep language model for software code,''
  \emph{CoRR}, vol. abs/1608.02715, 2016.

\bibitem{SANER18Dnn4C}
A.~T. Nguyen, T.~D. Nguyen, H.~D. Phan, and T.~N. Nguyen, ``A deep neural
  network language model with contexts for source code,'' in \emph{25th
  International Conference on Software Analysis, Evolution and Reengineering,
  {SANER} 2018, Campobasso, Italy, March 20-23, 2018}, 2018, pp. 323--334.

\bibitem{ICSE15GraLan}
A.~T. Nguyen and T.~N. Nguyen, ``Graph-based statistical language model for
  code,'' in \emph{37th {IEEE/ACM} International Conference on Software
  Engineering, {ICSE} 2015, Florence, Italy, May 16-24, 2015, Volume 1}, 2015,
  pp. 858--868.

\bibitem{ASE18RecRank}
X.~Liu, L.~Huang, and V.~Ng, ``Effective {API} recommendation without
  historical software repositories,'' in \emph{Proceedings of the 33rd
  {ACM/IEEE} International Conference on Automated Software Engineering, {ASE}
  2018, Montpellier, France, September 3-7, 2018}, 2018, pp. 282--292.

\bibitem{GGNN}
Y.~Li, D.~Tarlow, M.~Brockschmidt, and R.~S. Zemel, ``Gated graph sequence
  neural networks,'' \emph{CoRR}, vol. abs/1511.05493, 2015.

\bibitem{GNN}
F.~Scarselli, M.~Gori, A.~C. Tsoi, M.~Hagenbuchner, and G.~Monfardini, ``The
  graph neural network model,'' \emph{{IEEE} Trans. Neural Networks}, vol.~20,
  no.~1, pp. 61--80, 2009.

\bibitem{Almeida}
L.~B. Almeida, ``A learning rule for asynchronous perceptrons with feedback in
  a combinatorial environment.'' in \emph{Proceedings, 1st First International
  Conference on Neural Networks}, vol.~2, 1987, pp. 609--618.

\bibitem{Pineda}
F.~J. Pineda, ``Generalization of back-propagation to recurrent neural
  networks,'' \emph{Physical review letters}, vol.~59, no.~19, p. 2229, 1987.

\bibitem{GRU}
K.~Cho, B.~van Merrienboer, {\c{C}}.~G{\"{u}}l{\c{c}}ehre, F.~Bougares,
  H.~Schwenk, and Y.~Bengio, ``Learning phrase representations using {RNN}
  encoder-decoder for statistical machine translation,'' \emph{CoRR}, vol.
  abs/1406.1078, 2014.

\bibitem{camel_case}
\BIBentryALTinterwordspacing
``Camel case,'' 2020. [Online]. Available:
  \url{https://en.wikipedia.org/wiki/Camel\_case}
\BIBentrySTDinterwordspacing

\bibitem{glove}
\BIBentryALTinterwordspacing
``Glove,'' 2020. [Online]. Available:
  \url{https://nlp.stanford.edu/projects/glove}
\BIBentrySTDinterwordspacing

\bibitem{javaparser}
\BIBentryALTinterwordspacing
``Javaparser,'' 2020. [Online]. Available:
  \url{https://github.com/javaparser/javaparser/}
\BIBentrySTDinterwordspacing

\bibitem{Stanford_CoreNLP}
\BIBentryALTinterwordspacing
``Stanford corenlp,'' 2020. [Online]. Available:
  \url{https://stanfordnlp.github.io/CoreNLP/}
\BIBentrySTDinterwordspacing

\bibitem{tensorflow}
\BIBentryALTinterwordspacing
``Tensorflow,'' 2020. [Online]. Available:
  \url{https://github.com/tensorflow/tensorflow}
\BIBentrySTDinterwordspacing

\bibitem{source_code_of_ggnn}
\BIBentryALTinterwordspacing
``gg-nns reference implementation,'' 2020. [Online]. Available:
  \url{https://github.com/Microsoft/gated-graph-neural-network-samples}
\BIBentrySTDinterwordspacing

\bibitem{package}
\BIBentryALTinterwordspacing
``Replication package,'' 2020. [Online]. Available:
  \url{https://apireccst.wixsite.com/apirec-cst}
\BIBentrySTDinterwordspacing

\bibitem{DeepAPIRec}
C.~Chen, X.~Peng, J.~Sun, Z.~Xing, X.~Wang, Y.~Zhao, H.~Zhang, and W.~Zhao,
  ``Generative {API} usage code recommendation with parameter concretization,''
  \emph{{SCIENCE} {CHINA} Information Sciences}, vol.~62, no.~9, pp.
  192\,103:1--192\,103:22, 2019.

\bibitem{Groum}
T.~T. Nguyen, H.~A. Nguyen, N.~H. Pham, J.~M. Al{-}Kofahi, and T.~N. Nguyen,
  ``Graph-based mining of multiple object usage patterns,'' in
  \emph{Proceedings of the 7th joint meeting of the European Software
  Engineering Conference and the {ACM} {SIGSOFT} International Symposium on
  Foundations of Software Engineering, 2009, Amsterdam, The Netherlands, August
  24-28, 2009}, 2009, pp. 383--392.

\bibitem{Galaxy}
\BIBentryALTinterwordspacing
``Galaxy,'' 2020. [Online]. Available:
  \url{https://github.com/puniverse/galaxy}
\BIBentrySTDinterwordspacing

\bibitem{Log4j}
\BIBentryALTinterwordspacing
``Log4j,'' 2020. [Online]. Available: \url{https://github.com/apache/log4j}
\BIBentrySTDinterwordspacing

\bibitem{JGit}
\BIBentryALTinterwordspacing
``Jgit,'' 2020. [Online]. Available: \url{https://github.com/eclipse/jgit}
\BIBentrySTDinterwordspacing

\bibitem{Froyo-Email}
\BIBentryALTinterwordspacing
``Froyo-email,'' 2020. [Online]. Available:
  \url{https://github.com/Dustinmj/Froyo\_Email}
\BIBentrySTDinterwordspacing

\bibitem{Grid-Sphere}
\BIBentryALTinterwordspacing
``Grid-sphere,'' 2020. [Online]. Available:
  \url{https://github.com/brandt/GridSphere}
\BIBentrySTDinterwordspacing

\bibitem{Itext}
\BIBentryALTinterwordspacing
``Itext,'' 2020. [Online]. Available: \url{https://github.com/itext/itextpdf}
\BIBentrySTDinterwordspacing

\bibitem{FSE16APIREC}
A.~T. Nguyen, M.~Hilton, M.~Codoban, H.~A. Nguyen, L.~Mast, E.~Rademacher,
  T.~N. Nguyen, and D.~Dig, ``{API} code recommendation using statistical
  learning from fine-grained changes,'' in \emph{Proceedings of the 24th {ACM}
  {SIGSOFT} International Symposium on Foundations of Software Engineering,
  {FSE} 2016, Seattle, WA, USA, November 13-18, 2016}, 2016, pp. 511--522.

\bibitem{MRR}
V.~J. Hellendoorn, S.~Proksch, H.~C. Gall, and A.~Bacchelli, ``When code
  completion fails: a case study on real-world completions,'' in
  \emph{Proceedings of the 41st International Conference on Software
  Engineering, {ICSE} 2019, Montreal, QC, Canada, May 25-31, 2019}, 2019, pp.
  960--970.

\bibitem{t1}
\BIBentryALTinterwordspacing
``Stack overflow question,'' 2020. [Online]. Available:
  \url{https://stackoverflow.com/questions/326390/}
\BIBentrySTDinterwordspacing

\bibitem{t2}
\BIBentryALTinterwordspacing
``Stack overflow question,'' 2020. [Online]. Available:
  \url{https://stackoverflow.com/questions/223918/}
\BIBentrySTDinterwordspacing

\bibitem{t3}
\BIBentryALTinterwordspacing
``Stack overflow question,'' 2020. [Online]. Available:
  \url{https://stackoverflow.com/questions/415953/}
\BIBentrySTDinterwordspacing

\bibitem{t4}
\BIBentryALTinterwordspacing
``Stack overflow question,'' 2020. [Online]. Available:
  \url{https://stackoverflow.com/questions/160970/}
\BIBentrySTDinterwordspacing

\bibitem{t5}
\BIBentryALTinterwordspacing
``Stack overflow question,'' 2020. [Online]. Available:
  \url{https://stackoverflow.com/questions/1844688/}
\BIBentrySTDinterwordspacing

\bibitem{t6}
\BIBentryALTinterwordspacing
``Stack overflow question,'' 2020. [Online]. Available:
  \url{https://stackoverflow.com/questions/428918/}
\BIBentrySTDinterwordspacing

\bibitem{ICSM09David}
D.~M. Pletcher and D.~Hou, ``{BCC:} enhancing code completion for better {API}
  usability,'' in \emph{25th {IEEE} International Conference on Software
  Maintenance {(ICSM} 2009), September 20-26, 2009, Edmonton, Alberta, Canada},
  2009, pp. 393--394.

\bibitem{ICSE10Daqing}
D.~Hou and D.~M. Pletcher, ``Towards a better code completion system by {API}
  grouping, filtering, and popularity-based ranking,'' in \emph{Proceedings of
  the 2nd International Workshop on Recommendation Systems for Software
  Engineering, {RSSE} 2010, Cape Town, South Africa, May 4, 2010}, 2010, pp.
  26--30.

\bibitem{ICSM11Daqing}
------, ``An evaluation of the strategies of sorting, filtering, and grouping
  {API} methods for code completion,'' in \emph{{IEEE} 27th International
  Conference on Software Maintenance, {ICSM} 2011, Williamsburg, VA, USA,
  September 25-30, 2011}, 2011, pp. 233--242.

\bibitem{identifier-based}
L.~Heinemann, V.~Bauer, M.~Herrmannsdoerfer, and B.~Hummel, ``Identifier-based
  context-dependent {API} method recommendation,'' in \emph{16th European
  Conference on Software Maintenance and Reengineering, {CSMR} 2012, Szeged,
  Hungary, March 27-30, 2012}, 2012, pp. 31--40.

\bibitem{BMN}
M.~Bruch, M.~Monperrus, and M.~Mezini, ``Learning from examples to improve code
  completion systems,'' in \emph{Proceedings of the 7th joint meeting of the
  European Software Engineering Conference and the {ACM} {SIGSOFT}
  International Symposium on Foundations of Software Engineering, 2009,
  Amsterdam, The Netherlands, August 24-28, 2009}, 2009, pp. 213--222.

\bibitem{CSCC}
M.~Asaduzzaman, C.~K. Roy, K.~A. Schneider, and D.~Hou, ``A simple, efficient,
  context-sensitive approach for code completion,'' \emph{Journal of Software:
  Evolution and Process}, vol.~28, no.~7, pp. 512--541, 2016.

\bibitem{CoRR15APIMining}
J.~M. Fowkes and C.~A. Sutton, ``Parameter-free probabilistic {API} mining at
  github scale,'' \emph{CoRR}, vol. abs/1512.05558, 2015.

\bibitem{MSR13APIMing}
J.~Wang, Y.~Dang, H.~Zhang, K.~Chen, T.~Xie, and D.~Zhang, ``Mining succinct
  and high-coverage {API} usage patterns from source code,'' in
  \emph{Proceedings of the 10th Working Conference on Mining Software
  Repositories, {MSR} '13, San Francisco, CA, USA, May 18-19, 2013}, 2013, pp.
  319--328.

\bibitem{MSR06APIMining}
H.~Zhong, T.~Xie, L.~Zhang, J.~Pei, and H.~Mei, ``{MAPO:} mining and
  recommending {API} usage patterns,'' in \emph{{ECOOP} 2009 - Object-Oriented
  Programming, 23rd European Conference, Genoa, Italy, July 6-10, 2009.
  Proceedings}, 2009, pp. 318--343.

\bibitem{FSE09GrouMiner}
T.~T. Nguyen, H.~A. Nguyen, N.~H. Pham, J.~M. Al{-}Kofahi, and T.~N. Nguyen,
  ``Graph-based mining of multiple object usage patterns,'' in
  \emph{Proceedings of the 7th joint meeting of the European Software
  Engineering Conference and the {ACM} {SIGSOFT} International Symposium on
  Foundations of Software Engineering, 2009, Amsterdam, The Netherlands, August
  24-28, 2009}, 2009, pp. 383--392.

\bibitem{ICSE12Grapacc}
A.~T. Nguyen, T.~T. Nguyen, H.~A. Nguyen, A.~Tamrawi, H.~V. Nguyen, J.~M.
  Al{-}Kofahi, and T.~N. Nguyen, ``Graph-based pattern-oriented,
  context-sensitive source code completion,'' in \emph{34th International
  Conference on Software Engineering, {ICSE} 2012, June 2-9, 2012, Zurich,
  Switzerland}, 2012, pp. 69--79.

\bibitem{DL4SourceCode_1}
M.~Allamanis, H.~Peng, and C.~A. Sutton, ``A convolutional attention network
  for extreme summarization of source code,'' in \emph{Proceedings of the 33nd
  International Conference on Machine Learning, {ICML} 2016, New York City, NY,
  USA, June 19-24, 2016}, 2016, pp. 2091--2100.

\bibitem{CodeGeneration_1}
P.~Yin and G.~Neubig, ``A syntactic neural model for general-purpose code
  generation,'' in \emph{Proceedings of the 55th Annual Meeting of the
  Association for Computational Linguistics, {ACL} 2017, Vancouver, Canada,
  July 30 - August 4, Volume 1: Long Papers}, 2017, pp. 440--450.

\bibitem{CodeGeneration_2}
Z.~Sun, Q.~Zhu, L.~Mou, Y.~Xiong, G.~Li, and L.~Zhang, ``A grammar-based
  structural {CNN} decoder for code generation,'' \emph{CoRR}, vol.
  abs/1811.06837, 2018.

\bibitem{CodeGeneration_3}
W.~Ling, P.~Blunsom, E.~Grefenstette, K.~M. Hermann, T.~Kocisk{\'{y}}, F.~Wang,
  and A.~W. Senior, ``Latent predictor networks for code generation,'' in
  \emph{Proceedings of the 54th Annual Meeting of the Association for
  Computational Linguistics, {ACL} 2016, August 7-12, 2016, Berlin, Germany,
  Volume 1: Long Papers}, 2016.

\bibitem{CodeGeneration_4}
M.~Rabinovich, M.~Stern, and D.~Klein, ``Abstract syntax networks for code
  generation and semantic parsing,'' in \emph{Proceedings of the 55th Annual
  Meeting of the Association for Computational Linguistics, {ACL} 2017,
  Vancouver, Canada, July 30 - August 4, Volume 1: Long Papers}, 2017, pp.
  1139--1149.

\bibitem{CodeGeneration_5}
L.~Mou, R.~Men, G.~Li, L.~Zhang, and Z.~Jin, ``On end-to-end program generation
  from user intention by deep neural networks,'' \emph{CoRR}, vol.
  abs/1510.07211, 2015.

\bibitem{DeepAPILearning}
X.~Gu, H.~Zhang, D.~Zhang, and S.~Kim, ``Deep {API} learning,'' in
  \emph{Proceedings of the 24th {ACM} {SIGSOFT} International Symposium on
  Foundations of Software Engineering, {FSE} 2016, Seattle, WA, USA, November
  13-18, 2016}, 2016, pp. 631--642.

\bibitem{DeepCS}
X.~Gu, H.~Zhang, and S.~Kim, ``Deep code search,'' in \emph{Proceedings of the
  40th International Conference on Software Engineering, {ICSE} 2018,
  Gothenburg, Sweden, May 27 - June 03, 2018}, 2018, pp. 933--944.

\bibitem{CommentGeneration_1}
X.~Hu, G.~Li, X.~Xia, D.~Lo, and Z.~Jin, ``Deep code comment generation,'' in
  \emph{Proceedings of the 26th Conference on Program Comprehension, {ICPC}
  2018, Gothenburg, Sweden, May 27-28, 2018}, 2018, pp. 200--210.

\bibitem{CommentGeneration_2}
X.~Hu, G.~Li, X.~Xia, D.~Lo, S.~Lu, and Z.~Jin, ``Summarizing source code with
  transferred {API} knowledge,'' in \emph{Proceedings of the Twenty-Seventh
  International Joint Conference on Artificial Intelligence, {IJCAI} 2018, July
  13-19, 2018, Stockholm, Sweden.}, 2018, pp. 2269--2275.

\bibitem{CommentGeneration_3}
Y.~Liang and K.~Q. Zhu, ``Automatic generation of text descriptive comments for
  code blocks,'' in \emph{Proceedings of the Thirty-Second {AAAI} Conference on
  Artificial Intelligence, (AAAI-18), the 30th innovative Applications of
  Artificial Intelligence (IAAI-18), and the 8th {AAAI} Symposium on
  Educational Advances in Artificial Intelligence (EAAI-18), New Orleans,
  Louisiana, USA, February 2-7, 2018}, 2018, pp. 5229--5236.

\bibitem{CommentGeneration_4}
Y.~Wan, Z.~Zhao, M.~Yang, G.~Xu, H.~Ying, J.~Wu, and P.~S. Yu, ``Improving
  automatic source code summarization via deep reinforcement learning,'' in
  \emph{Proceedings of the 33rd {ACM/IEEE} International Conference on
  Automated Software Engineering, {ASE} 2018, Montpellier, France, September
  3-7, 2018}, 2018, pp. 397--407.

\bibitem{CommentGeneration_5}
S.~Iyer, I.~Konstas, A.~Cheung, and L.~Zettlemoyer, ``Summarizing source code
  using a neural attention model,'' in \emph{Proceedings of the 54th Annual
  Meeting of the Association for Computational Linguistics, {ACL} 2016, August
  7-12, 2016, Berlin, Germany, Volume 1: Long Papers}, 2016.

\bibitem{DefectPrediction_1}
X.~Yang, D.~Lo, X.~Xia, Y.~Zhang, and J.~Sun, ``Deep learning for just-in-time
  defect prediction,'' in \emph{2015 {IEEE} International Conference on
  Software Quality, Reliability and Security, {QRS} 2015, Vancouver, BC,
  Canada, August 3-5, 2015}, 2015, pp. 17--26.

\bibitem{DefectPrediction_2}
S.~Wang, T.~Liu, and L.~Tan, ``Automatically learning semantic features for
  defect prediction,'' in \emph{Proceedings of the 38th International
  Conference on Software Engineering, {ICSE} 2016, Austin, TX, USA, May 14-22,
  2016}, 2016, pp. 297--308.

\end{thebibliography}
